\documentclass[12pt]{article}
\usepackage{latexsym}
\usepackage{amsmath}
\usepackage{amssymb}
\usepackage{epsf}
\usepackage{epsfig}
\usepackage{cite}
\usepackage{accents}
\usepackage{booktabs}
\usepackage{lineno}

\newcommand{\lsim}{
\mathrel{\hbox{\rlap{\hbox{\lower4pt\hbox{$\sim$}}}\hbox{$<$}}}}

\newcommand{\gsim}{
\mathrel{\hbox{\rlap{\hbox{\lower4pt\hbox{$\sim$}}}\hbox{$>$}}}}

\newcommand*{\fancybar}{\scalebox{.4}{(}\raisebox{-1.7pt}{\bf --}\scalebox{.4}{)}}
\newcommand*{\brabar}[1]{\accentset{\fancybar}{#1}}

\def\D0{D\O }

\begin{document}

\begin{titlepage}

\vspace*{-0.0truecm}

\begin{flushright}
Nikhef-2012-012
\end{flushright}

\vspace*{1.3truecm}

\begin{center}
\boldmath
{\Large{\bf Exploring $B_s \to D_s^{(*)\pm} K^\mp$ Decays in the\\ 
Presence of a Sizable Width Difference $\Delta\Gamma_s$}}
\unboldmath
\end{center}

\vspace{0.9truecm}

\begin{center}
{\bf Kristof De Bruyn\,${}^a$, Robert Fleischer\,${}^{a,b}$, Robert Knegjens\,${}^a$, \\
Marcel Merk\,${}^{a,b}$, Manuel Schiller\,${}^a$ and
Niels Tuning\,${}^a$}

\vspace{0.5truecm}

${}^a${\sl Nikhef, Science Park 105, NL-1098 XG Amsterdam, Netherlands}

${}^b${\sl  Department of Physics and Astronomy, Vrije Universiteit Amsterdam,\\
NL-1081 HV Amsterdam, Netherlands}

\end{center}

\vspace{1.4cm}
\begin{abstract}
\vspace{0.2cm}\noindent
The $B_s \to D_s^{(*)\pm} K^\mp$ decays allow a theoretically clean determination 
of $\phi_s+\gamma$, where $\phi_s$ is the $B^0_s$--$\bar B^0_s$ mixing phase and 
$\gamma$ the usual angle of the unitarity triangle. A sizable $B_s$ decay width 
difference $\Delta\Gamma_s$ was recently established, which leads to subtleties in 
analyses of the $B_s \to D_s^{(*)\pm} K^\mp$ branching ratios but also offers new 
``untagged'' observables, which do not require a distinction between initially present 
$B^0_s$ or $\bar B^0_s$ mesons. We clarify these effects and address recent 
measurements of the ratio of the $B_s\to D_s^\pm K^\mp$, $B_s\to D_s^\pm\pi^\mp$ 
branching ratios. In anticipation of future LHCb analyses, we apply the $SU(3)$
flavour symmetry of strong interactions to convert the $B$-factory data for 
$B_d\to D^{(*)\pm}\pi^\mp$,  $B_d\to D_s^{\pm}\pi^\mp$ decays into predictions of the 
$B_s \to D_s^{(*)\pm} K^\mp$ observables, and discuss strategies for
the extraction of $\phi_s+\gamma$, with a special focus on untagged observables and the 
resolution of discrete ambiguities. Using our theoretical predictions as a guideline, we 
make simulations to estimate experimental sensitivities, and extrapolate to the end of the 
planned LHCb upgrade. We find that the interplay between the untagged observables,
which are accessible thanks to the sizable $\Delta\Gamma_s$, and the mixing-induced 
CP asymmetries, which require tagging, will play the key role for the experimental
determination of $\phi_s+\gamma$. 
\end{abstract}

\vspace*{0.5truecm}
\vfill
\noindent
August 2012
\vspace*{0.5truecm}

\end{titlepage}

\thispagestyle{empty}
\vbox{}
\newpage

\setcounter{page}{1}

\section{Introduction} \label{sec:intro}
The decays $B_s \to D_s^{(*)\pm} K^\mp$ only receive contributions from 
tree-diagram-like topologies\footnote{We use the notation $B_s = \brabar{B}_s^0$.}. 
Since both $B^0_s$ and $\bar B^0_s$ mesons
can decay into the $D_s^{(*)\pm} K^\mp$ final states, interference effects between
$B^0_s$--$\bar B^0_s$ mixing and decay processes allow a theoretically clean
determination of the CP-violating phase $\phi_s+\gamma$ 
\cite{BsDsK,RF-BsDsK}, where $\phi_s$ is
the $B^0_s$--$\bar B^0_s$ mixing phase and $\gamma$ the corresponding angle
of the unitarity triangle. As $\phi_s$ can be extracted separately, with the latest
experimental average given by \cite{HFAG} 
\begin{equation}\label{phis-average}
	\phi_s=\left(-2.5^{+5.2}_{-4.9}\right)^\circ,
\end{equation}
$\gamma$ can be determined. 

The central question is then whether this value will agree with 
$\gamma$ determinations from decays with penguin contributions, such as the
$B^0_d\to \pi^+\pi^-$, $B^0_s\to K^+K^-$ system \cite{RF-BsKK}. The current picture
of direct determinations of $\gamma$ from tree decays can be summarized as follows:
\begin{equation}
\gamma=\left\{\begin{array}{ll}
(66\pm12)^\circ & \mbox{(CKMfitter Collaboration \cite{CKMfitter})}\\
(76\pm10)^\circ & (\mbox{UTfit Collaboration \cite{UTfit}}).
\end{array}
 \right.
\end{equation}
On the other hand, a recent analysis of the $B^0_d\to \pi^+\pi^-$, $B^0_s\to K^+K^-$ 
system gives 
\begin{equation}\label{gam-det}
\gamma=(68\pm7)^\circ,
\end{equation}
where the error also takes $SU(3)$-breaking corrections into account \cite{Fleischer:2010ib}.

In the present paper, we assume that the relevant decay amplitudes are described 
by the Standard Model (SM).  Applying the formalism developed in Ref.~\cite{RF-BsDsK}, 
we shall explore the $B_s \to D_s^{(*)\pm} K^\mp$ channels both in view of recent 
experimental developments and measurements to be performed by the 
LHCb collaboration in this decade.

Using the $B^0_s \to J/\psi \phi$ channel, the LHCb experiment has recently established a 
non-vanishing decay width difference of the $B_s$-meson system, which is characterized 
by the following parameter \cite{LHCb-Mor-12}:
\begin{equation}\label{ys-LHCb}
	y_s \equiv \frac{\Delta\Gamma_s}{2\,\Gamma_s}\equiv
	\frac{\Gamma_{\rm L}^{(s)} - \Gamma_{\rm H}^{(s)}}{2\,\Gamma_s}= 0.088 \pm 0.014.
\end{equation}
Here $\tau_{B_s}^{-1} \equiv \Gamma_s \equiv \bigl[\Gamma_{\rm L}^{(s)} + 
\Gamma_{\rm H}^{(s)}\bigr]/2=\left(0.6580 \pm 0.0085\right)\mbox{ps}^{-1}$ \cite{LHCb-Mor-12}
denotes the inverse of the $B_s$ mean lifetime $\tau_{B_s}$. A discrete ambiguity
could also be resolved \cite{LHCb-DGambig}, thereby leaving us with the sign of
$\Delta\Gamma_s$ in (\ref{ys-LHCb}), which is in agreement with the SM expectation 
(for a recent review, see \cite{lenz}). 

This new development in the exploration of the $B_s$-meson system has important 
consequences:
\begin{itemize}
\item Untagged $B_s$ decay data samples, where no distinction is made between initially, i.e.\
at time $t=0$, present $B^0_s$ or $\bar B^0_s$ mesons, allow for an extraction of interesting
observables \cite{RF-BsDsK}.
\item A subtle difference arises between the branching ratios extracted experimentally,
and those usually considered by theory~\cite{BR-paper}.
\end{itemize}

First measurements of the  $B_s \to D_s^\pm K^\mp$ branching ratios are available
from the CDF \cite{CDF-BsDsK}, Belle \cite{Belle-BsDsK}
and LHCb \cite{LHCb-BsDsK} collaborations:
\begin{equation}\label{LHCb-CDF-BR}
	\frac{\text{BR}(B_s\rightarrow 
	D_s^\pm K^\mp)_{\rm exp}}{\text{BR}(B_s\rightarrow D_s^\pm \pi^\mp)_{\rm exp}} = \left\{
	\begin{array}{lcl}	
		0.097 \pm 0.018\:(\text{stat.}) \pm 0.009\:(\text{syst.}) &  &{\rm [CDF],}\\
		0.065^{+0.035}_{-0.029} \:(\text{stat.}) &  &{\rm [Belle],}\\
		0.0646 \pm 0.0043\:(\text{stat.}) \pm 0.0025\:(\text{syst.}) &  & 
		{\rm [LHCb];}
	\end{array}\right. 
\end{equation}
the errors of the Belle result are dominated by the small $B_s\rightarrow D_s^\pm K^\mp$ 
data sample. 
We shall clarify the impact of $\Delta\Gamma_s$ on this ratio of CP-averaged experimental
branching ratios and convert the experimental numbers into constraints on the 
hadronic parameter characterizing the interference effects discussed above. 

As was pointed out in Ref.~\cite{RF-BsDsK}, the observables of the 
$B_s \to D_s^{(*)\pm} K^\mp$ channels can be related to those of the
$B_d \to D^{(*)\pm} \pi^\mp$ decays through the $U$-spin symmetry of 
strong interactions. We shall use $B$-factory data for the latter decays
obtained by the BaBar and Belle collaborations, with further constraints from
$B_d\to D_s^\pm\pi^\mp$ modes, to make predictions for the
$B_s \to D_s^{(*)\pm} K^\mp$ observables that will serve as a guideline for the
expected experimental picture. In this analysis, we specifically find that -- thanks to the
sizable value of $\Delta\Gamma_s$ -- untagged data samples of $B_s \to D_s^{(*)\pm} K^\mp$
decays can be efficiently combined with mixing-induced CP asymmetries of tagged
analyses to extract $\phi_s+\gamma$ in an unambiguous way. 

The outline is as follows: in Section~\ref{sec:untagged}, we discuss untagged measurements
of the  $B_s \to D_s^{(*)\pm} K^\mp$ decays and their effective lifetimes, addressing
also the results listed in (\ref{LHCb-CDF-BR}). In Section~\ref{sec:Bd},
we apply $SU(3)$ flavour symmetry to extract the hadronic parameters
characterizing the $B_s \to D_s^{(*)\pm} K^\mp$ decays from the $B$-factory data 
for the $B_d\to D^{(*)\pm} \pi^\mp$ and $B_d\to D_s^\pm\pi^\mp$ channels. 
In Section~\ref{sec:CP}, we discuss the
extraction of $\phi_s+\gamma$ from the tagged and untagged $B_s \to D_s^{(*)\pm} K^\mp$ 
observables, with a special emphasis on resolving the discrete ambiguities. 
The hadronic parameters obtained in Section~\ref{sec:Bd} are used in Section~\ref{sec:exp}
to predict the relevant $B_s \to D_s^{(*)\pm} K^\mp$ observables, which then serve as 
an input for exploring the experimental prospects. Finally, we summarize our 
conclusions in Section~\ref{sec:concl}.

\boldmath
\section{Untagged Observables}\label{sec:untagged}
\unboldmath
The time-dependent, untagged $B_s \to D_s^{(*)+} K^-$ decay rates can be written
as follows \cite{RF-BsDsK}:
\begin{align}
	\langle \Gamma(B_s(t)\to D_s^{(*)+} K^-)\rangle &\equiv 
	\Gamma(B^0_s(t)\to D_s^{(*)+} K^-) + \Gamma(\bar{B}^0_s(t)\to D_s^{(*)+} K^-) \notag\\
	&= R\, e^{-t/\tau_{B_s}}\left[\cosh\left(y_s\frac{t}{\tau_{B_s}}\right) + 
	{\cal A}_{\Delta\Gamma}\sinh\left(y_s\frac{t}{\tau_{B_s}}\right) \right],
	\label{untagged}
\end{align}
where 
\begin{equation}\label{R-def}
R \equiv \langle \left.\Gamma(B_s(t)\to D_s^{(*)+} K^-)\rangle\right|_{t=0} =
	\left.\langle \Gamma(B_s(t)\to D_s^{(*)-} K^+) \rangle\right|_{t=0}.
\end{equation}
The time-dependent, untagged $B_s \to D_s^{(*)-} K^+$ rate into the CP-conjugate final state 
can be straightforwardly obtained from (\ref{untagged}) by replacing ${\cal A}_{\Delta\Gamma}$ with 
$\overline{\cal A}_{\Delta\Gamma}$. The latter observables take the form
\begin{equation}\label{ADG-expr}
	{\cal A}_{\Delta\Gamma} = 
	- (-1)^L \frac{2\,x_s}{1+x_s^2} \cos(\phi_s + \gamma + \delta_s), \quad
	\overline{\cal A}_{\Delta\Gamma} = 
	- (-1)^L \frac{2\,x_s}{1+x_s^2} \cos(\phi_s + \gamma - \delta_s),
\end{equation}
where $L$ denotes the angular momentum of the final state\footnote{For simplicity, we did not
introduce a label to distinguish between $D_s^{+} K^-$ and $ D_s^{*+} K^-$.}, the hadronic
parameter $x_s\propto R_b$ 
quantifies the strength of the
interference effects between the $B^0_s\to D_s^{(*)+} K^-$ and $\bar B^0_s\to D_s^{(*)+} K^-$
decay processes induced through $B^0_s$--$\bar B^0_s$ mixing, and $\delta_s$ is an 
associated CP-conserving strong phase difference \cite{RF-BsDsK};
the parameter $R_b\propto |V_{ub}/(\lambda V_{cb})|\sim 0.4$ measures one side of the unitarity 
triangle.

The branching ratios of $B_s$ decays are determined experimentally as 
time-integrated untagged rates \cite{BR-paper} (see also Eqs. (21) and (22) in Ref.~\cite{DFN}):
\begin{equation}
	{\rm BR}(B_s\to D_s^{(*)\pm} K^\mp)_{\rm exp} \equiv \frac{1}{2} \int_0^\infty 
	\langle \Gamma(B_s\to D_s^{(*)\pm} K^\mp)\rangle\,dt.
\end{equation}
On the other hand, the branching ratio corresponding to the untagged rate
at $t=0$, where $B^0_s$--$\bar B^0_s$ mixing is ``switched off", is usually considered by 
theorists. The conversion between this theoretical branching ratio and the experimental
branching ratio is given as follows \cite{BR-paper,DFN}:
\begin{equation}
	{\rm BR}(B_s\to D_s^{(*)+} K^-)_{\rm theo} = 
	\left[\frac{1-y_s^2}{1+{\cal A}_{\Delta\Gamma}\,y_s}\right]
	{\rm BR}(B_s\to D_s^{(*)+} K^-)_{\rm exp},
	\label{expTheoConv}
\end{equation}
where an analogous expression involving $\overline{\cal A}_{\Delta\Gamma}$ holds 
for the $D_s^{(*)-} K^+$ final states. It is interesting to note that we have
\begin{equation}
{\rm BR}(B_s\to D_s^{(*)+} K^-)_{\rm theo} = {\rm BR}(B_s\to D_s^{(*)-} K^+)_{\rm theo}
\end{equation}
thanks to (\ref{R-def}), which implies 
\begin{equation}\label{BR-OE}
\frac{{\rm BR}(B_s\to D_s^{(*)+} K^-)_{\rm exp}}{{\rm BR}(B_s\to D_s^{(*)-} K^+)_{\rm exp}}=
\frac{1+{\cal A}_{\Delta\Gamma}\,y_s}{1+ \overline{\cal A}_{\Delta\Gamma}\,y_s}.
\end{equation} 
Consequently, an established difference between the experimental $B_s \to D_s^{(*)-} K^+$
and $B_s \to D_s^{(*)+} K^-$ branching ratios would imply a difference between the 
${\cal A}_{\Delta\Gamma}$ and $\overline{\cal A}_{\Delta\Gamma}$ observables 
(see also Ref.~\cite{NandiNierste_DsK}):
\begin{equation}\label{BR-diff}
\frac{{\rm BR}(B_s\to D_s^{(*)+} K^-)_{\rm exp}-
{\rm BR}(B_s\to D_s^{(*)-} K^+)_{\rm exp}}{{\rm BR}(B_s\to D_s^{(*)+} K^-)_{\rm exp}+
{\rm BR}(B_s\to D_s^{(*)-} K^+)_{\rm exp}}=y_s\left[\frac{{\cal A}_{\Delta\Gamma}-
\overline{\cal A}_{\Delta\Gamma}}{2+y_s({\cal A}_{\Delta\Gamma}+
\overline{\cal A}_{\Delta\Gamma})}\right].
\end{equation}

In order to relate theory to experiment beyond an accuracy corresponding to the size 
of $y_s\sim 0.1$, we need theoretical input to determine ${\cal A}_{\Delta\Gamma}$ and
$\overline{\cal A}_{\Delta\Gamma}$. In Section~\ref{sec:Bd}, we will see that this results in 
large uncertainties for these observables. However, this input can be avoided with the help of the
effective decay lifetimes \cite{BR-paper}, defined as 

\begin{equation}\label{Teff}
	{\tau}_{\rm eff} \equiv 
	\frac{\int_0^{\infty}t\:\langle{\Gamma}
	(B_s\to D_s^{(*)+} K^-)\rangle dt}{\int_0^{\infty}\langle{\Gamma}(B_s\to D_s^{(*)+} 
	K^-)\rangle dt} 
	= \frac{\tau_{B_s}}{1- y_s^2}\left[\frac{1 +2\, {\cal A}_{\Delta\Gamma}\,y_s + y_s^2}
		{1 + {\cal A}_{\Delta\Gamma}\,y_s} \right],
\end{equation}
with an analogous expression for the  lifetimes $\overline{\tau}_{\rm eff}$  of the 
CP-conjugate $D_s^{(*)-} K^+$ final states (see also Ref.~\cite{NandiNierste_DsK}). 
We then obtain

\begin{equation}
	{\rm BR}(B_s\to D_s^{(*)+} K^-)_{\rm theo} = 
	\left[2 - (1-y_s^2)\,\tau_{\rm eff} \right]{\rm BR}(B_s\to D_s^{(*)+} K^-)_{\rm exp},
	\label{expTheoConvLifetime}	
\end{equation}
and correspondingly for the $D_s^{(*)-} K^+$ final states. These general relations hold also
should the $B_s \to D_s^{(*)\pm} K^\mp$ decay amplitudes receive contributions from
physics beyond the SM, which is not a plausible scenario.

Let us now have a closer look at the ratio (\ref{LHCb-CDF-BR}). Since the $B^0_s\to D_s^-\pi^+$,
$\bar B^0_s\to D_s^+\pi^-$ decays are flavour-specific, their ${\cal A}_{\Delta\Gamma}$, $\overline{\cal A}_{\Delta\Gamma}$
observables vanish. The branching ratios entering (\ref{LHCb-CDF-BR}) are  
averages of the experimental branching ratios over the final states:
\begin{equation}
{\rm BR}(B_s\to D_s^\pm K^\mp)_{\rm exp} 
\equiv \frac{1}{2}\left[ {\rm BR}(B_s\to D_s^{+} K^-)_{\rm exp}+
{\rm BR}(B_s\to D_s^{-} K^+)_{\rm exp} \right],
\end{equation}
with an analogous expression for ${\rm BR}(B_s\to D_s^\pm \pi^\mp)_{\rm exp}$.
Using \eqref{R-def} and its $B_s\to D_s^\pm \pi^\mp$ counterpart yields
\begin{equation}
	\frac{ {\rm BR}(B_s\to D_s^{(*)\pm} K^\mp)_{\rm exp} }
	{ {\rm BR}(B_s\to D_s^{(*)\pm} \pi^\mp)_{\rm exp}}
	= \left[1 + y_s \left(\frac{{\cal A}_{\Delta\Gamma} + 
	\overline{\cal A}_{\Delta\Gamma}}{2}\right)\right]\frac
	{ {\rm BR}(B_s\to D_s^{(*)\pm} K^\mp)_{\rm theo} }
	{ {\rm BR}(B_s\to D_s^{(*)\pm} \pi^\mp)_{\rm theo}}.
	\label{expTheoConvRatio}
\end{equation}

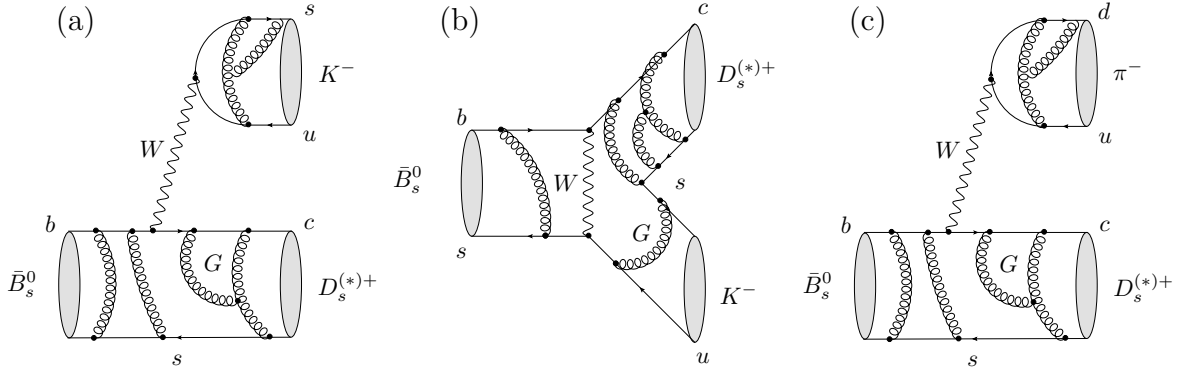
\begin{figure}[!t]
  \begin{center}
    \begin{picture}(250,140)(0,0)
      \put( -90,0){\includegraphics[scale=0.25]{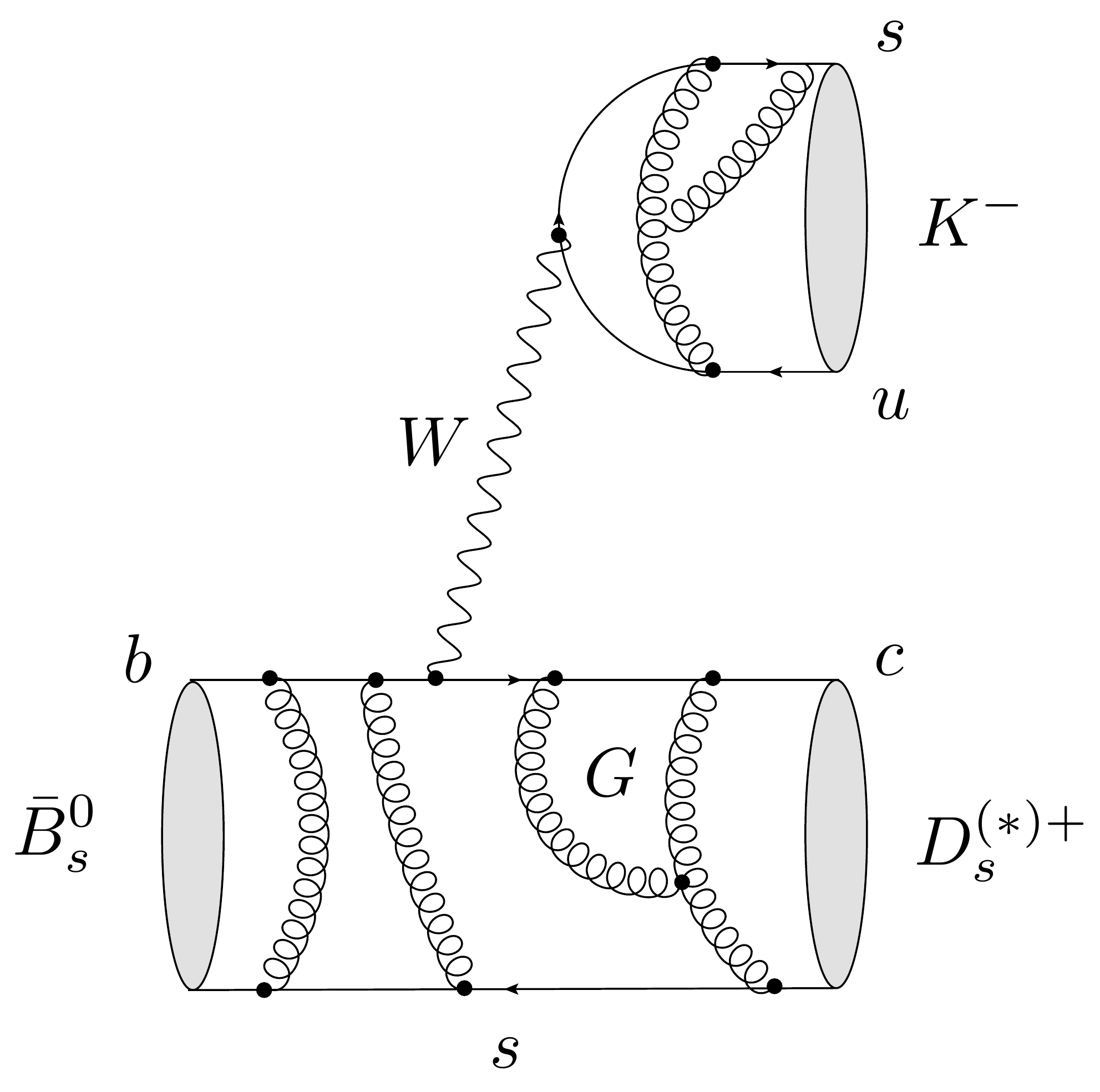}}
      \put( -70,130){(a)}
      \put( 55,0){\includegraphics[scale=0.25]{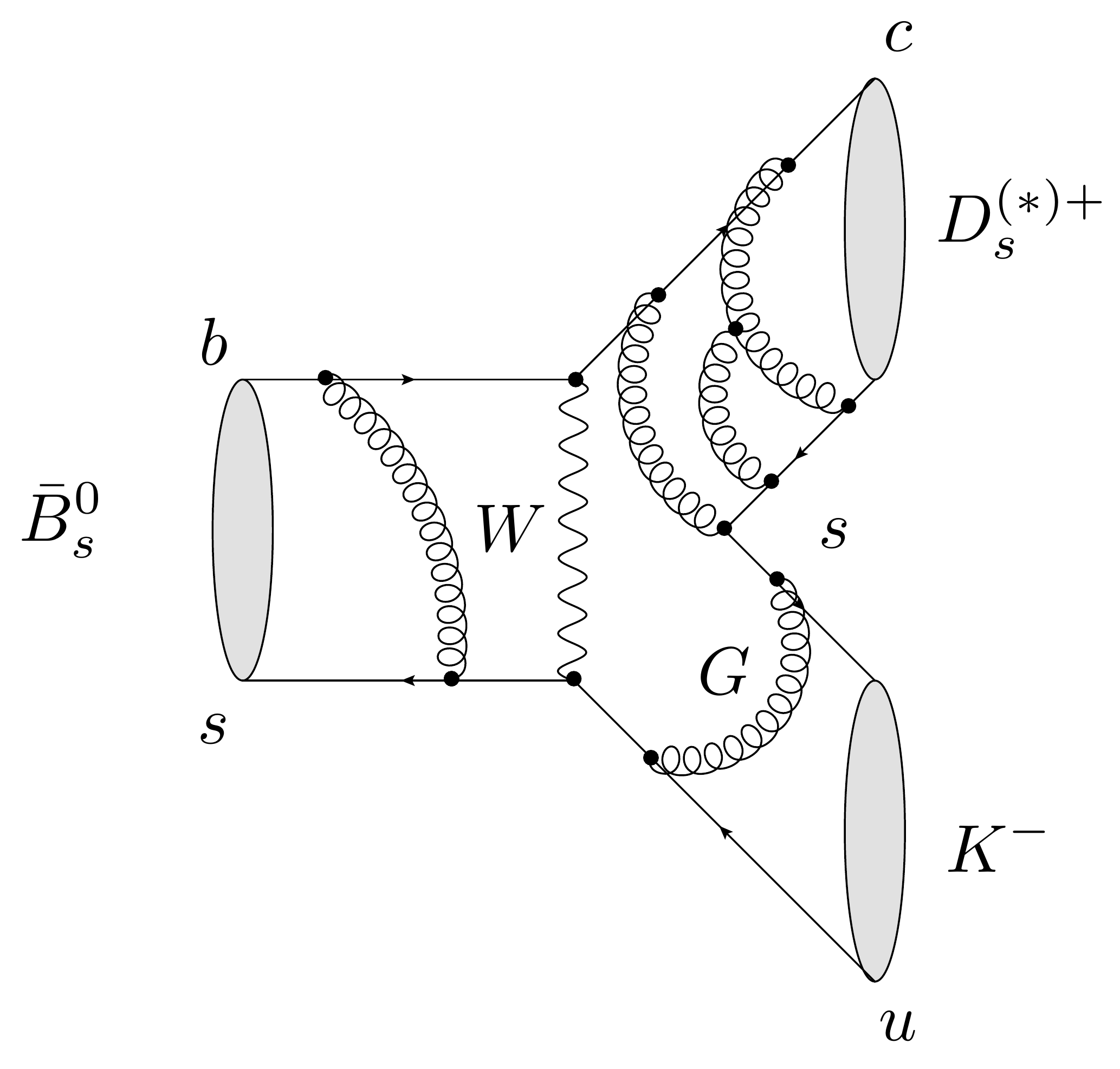}}
      \put( 75,130){(b)}
      \put(210,0){\includegraphics[scale=0.25]{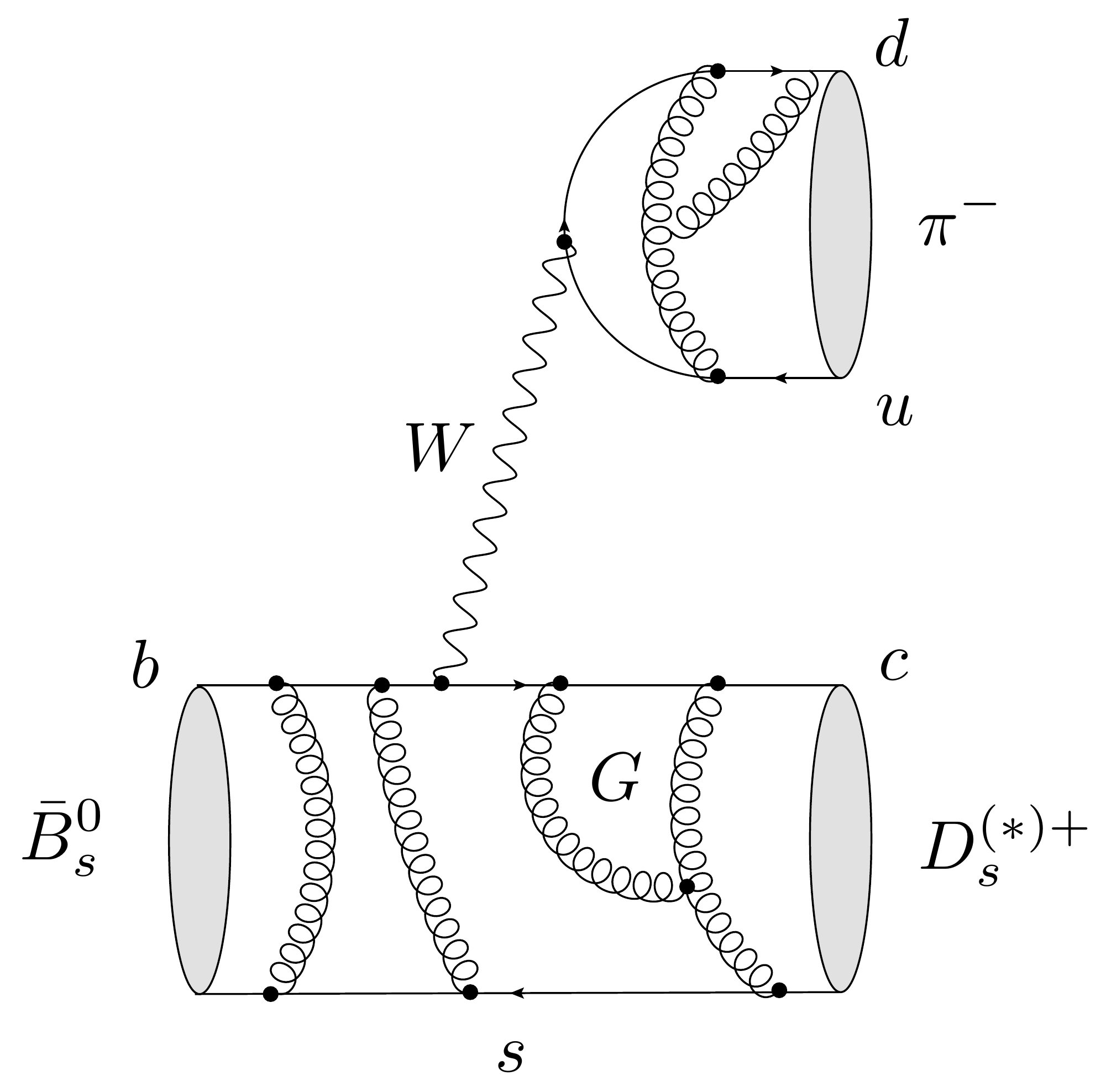}}
      \put(230,130){(c)}
    \end{picture}
    \caption[TCE]{The colour-allowed tree (a) and exchange (b) topologies contributing to the
     $\bar B^0_s\to D_s^{(*)+}K^-$ decay in comparison with the colour-allowed tree (c) topology
     of the $SU(3)$-related $\bar B^0_s\to D_s^{(*)+}\pi^-$ channel which does not receive
     exchange contributions because of the flavour content of its final state.}
    \label{fig:topol}
  \end{center}
\end{figure}

The ``factorization" of hadronic matrix elements is expected to work well for the amplitudes
of the $\bar B^0_s\to D_s^{(*)+}K^-$ and  $\bar B^0_s\to D_s^{(*)+}\pi^-$ decays 
\cite{fact,BGR,bjor,DG,BBNS,SCET}, which is also supported by 
experimental data \cite{fact-test}. In Fig.~\ref{fig:topol}, we illustrate the decay topologies 
characterizing these decays. Using the $SU(3)$ flavour symmetry to relate the 
$\bar B^0_s\to D_s^{(*)+}K^-$ amplitude to that of the $\bar B^0_s\to D_s^{(*)+}\pi^-$ 
channel (and correspondingly for the CP-conjugate processes), the ratio of the theoretical 
branching ratios in (\ref{expTheoConvRatio}) allows the extraction of the hadronic 
parameter $x_s$, as discussed in detail in Ref.~\cite{RF-BsDsK}:
\begin{equation}\label{xs-det-1}
	x_s=\sqrt{\left[\frac{{\cal C}^{(*)}}{\epsilon}\right]
	\left[\frac{{\rm BR}(B_s\to D_s^{(*)\pm} K^\mp)_{\rm theo} }
	{ {\rm BR}(B_s\to D_s^{(*)\pm} \pi^\mp)_{\rm theo}}\right] - 1}.
\end{equation}
Here 
\begin{equation}
\epsilon\equiv\frac{\lambda^2}{1-\lambda^2}=0.0534 \pm 0.0005
\end{equation}
involves the Wolfenstein parameter $\lambda\equiv|V_{us}|=0.2252\pm0.0009$ \cite{PDG}, while
the ${\cal C}^{(*)}$ coefficient can be written in the following form:
\begin{equation}\label{C-def}
{\cal C}^{(*)}\equiv\frac{\Phi_{D^{(*)}_s\pi}}{\Phi_{D^{(*)}_sK}}{\cal N}^{(*)}_F{\cal N}^{(*)}_a 
{\cal N}^{(*)}_E,
\end{equation}
where the $\Phi$ are straightforwardly calculable phase-space factors, and
\begin{equation}\label{NF}
{\cal N}^{(*)}_F \equiv \left[\frac{f_\pi}{f_K}
	\frac{F_{B_s\to D^{(*)}_s}(M_\pi^2)}{F_{B_s\to D^{(*)}_s}(M_K^2)}\right]^2
\end{equation}
describes factorizable $SU(3)$-breaking corrections through the ratios of
decay constants $f_K/f_\pi=1.197\pm 0.006$~\cite{PDG} and form 
factors.\footnote{For the calculation of the form-factor ratio in (\ref{NF}) 
we have assumed that the $q^2$ dependence is identical to that for $B_d\to D^{(*)-}\ell\nu$ 
decays~\cite{Caprini}.} On the other hand, the non-factorizable 
$SU(3)$-breaking corrections affecting the
ratio of the colour-allowed tree amplitudes governing the $\bar B^0_s\to D_s^{(*)+}K^-$ and
 $\bar B^0_s\to D_s^{(*)+}\pi^-$ channels are described by 
 \begin{equation}
 {\cal N}^{(*)}_a \equiv \left|\frac{a_1(D^{(*)}_s\pi)}{a_1(D^{(*)}_sK)}\right|^2.
 \end{equation}
 Finally, ${\cal N}^{(*)}_E$ takes into account that the $\bar B^0_s\to D_s^{(*)+}K^-$ 
 decays receive also contributions from exchange topologies, which have no counterparts
 in the $\bar B^0_s\to D_s^{(*)+}\pi^-$ processes, as can be seen in Fig.~\ref{fig:topol}:
 \begin{equation}\label{NE-def}
 {\cal N}^{(*)}_E\equiv\left|\frac{T_{ D_s^{(*)+}K^-}}{T_{ D_s^{(*)+}K^-}+E_{ D_s^{(*)+}K^-}}\right|^2.
 \end{equation}
Following the phenomenological analysis of Ref.~\cite{fact-test} using experimental data
to make factorization tests and to constrain the exchange topologies, we
find ${\cal N}^{(*)}_a\sim 1.00\pm0.02$ and ${\cal N}^{(*)}_E\sim 0.97\pm0.08$. The exchange
contributions can be probed further in the future through the $\bar B^0_s\to D^{(*)+}\pi^-$
channel, which receives only contributions from such topologies \cite{RF-BsDsK}.
Finally, we obtain the numerical value 
\begin{equation}
	{\cal C}^{(*)}  = 0.67 \pm 0.05.
\end{equation}

\begin{figure}[tbp] 
   \centering
	   \includegraphics[width=8.4truecm]{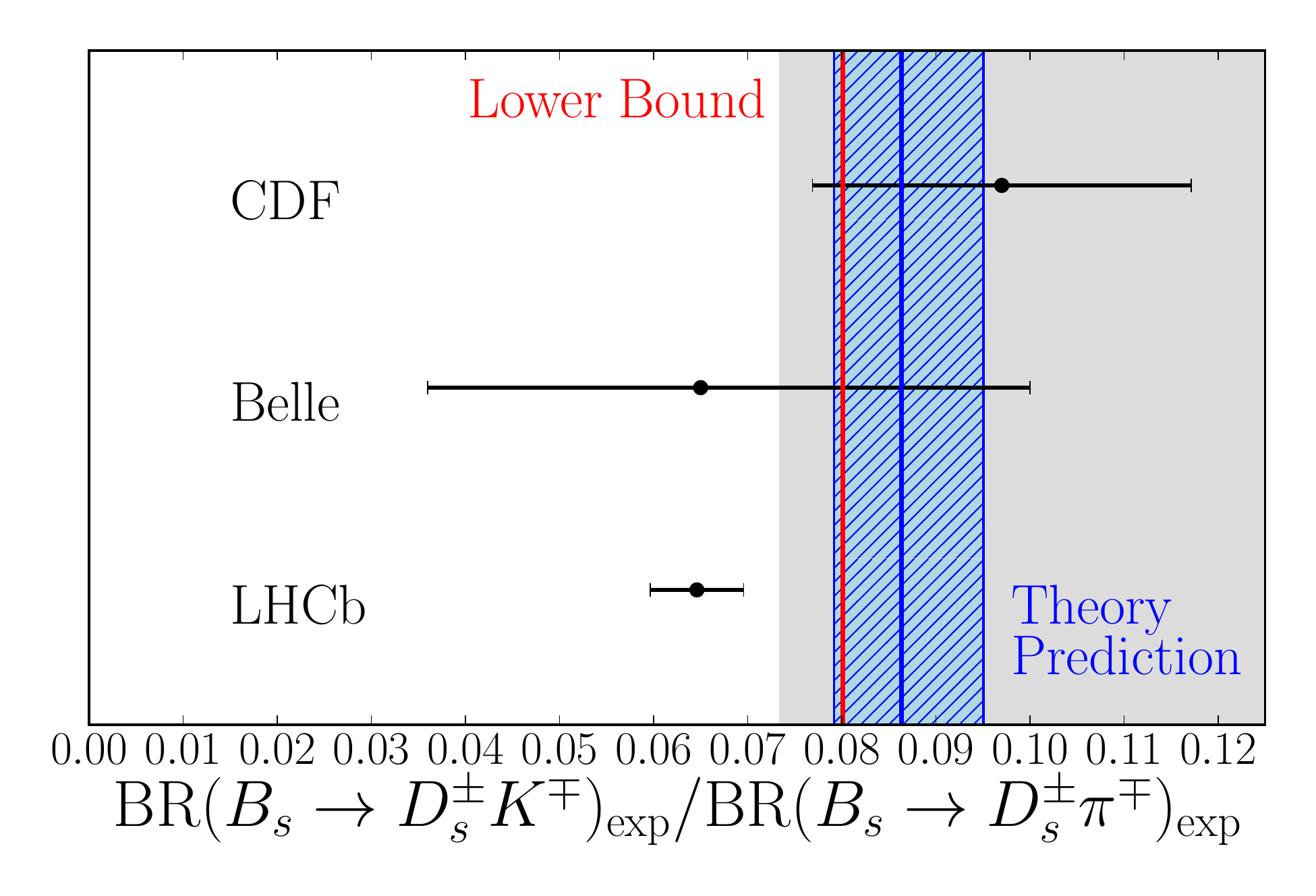}
	   \caption{Compilation of measurements of the ratio of branching
	    ratios as given in (\ref{LHCb-CDF-BR})
	     and comparison with the lower bound in (\ref{BR-bound}).
	     The theoretical prediction indicated by the vertical band 
	     corresponds to (\ref{eq:theo_predBR})
	     as given in Section~\ref{sec:exp}.}
   \label{fig:BRratio}
\end{figure}

Using now (\ref{ADG-expr}) and (\ref{expTheoConvRatio}), we arrive at
\begin{eqnarray}
\lefteqn{x_s = 	 y_s\, \cos\delta_s\,\cos(\phi_s+\gamma)}\nonumber\\
&& \pm \sqrt{ 
		\left[\frac{ {\cal C}^{(*)}}{\epsilon}\right]\left[
		\frac{ {\rm BR}(B_s\to D_s^{(*)\pm} K^\mp)_{\rm exp} }
	{ {\rm BR}(B_s\to D_s^{(*)\pm} \pi^\mp)_{\rm exp}}\right]
	 -1 + y_s^2\, \cos^2\delta_s\,\cos^2(\phi_s+\gamma)},\label{xs-BR-det}
\end{eqnarray}
where $x_s$ was defined as a positive parameter \cite{RF-BsDsK}. For the numerical
values of $\phi_s$ and $\gamma$ in (\ref{phis-average}) and (\ref{gam-det}), respectively,
the CDF result in (\ref{LHCb-CDF-BR}) gives
$x_s=0.46 \pm 0.27\:(\text{BR}) \pm 0.11\:(\mathcal{C}) \pm 0.04\:(\delta_s)$. 
This value for $x_s$ is consistent with theoretical expectations  \cite{RF-BsDsK} and the picture 
discussed in the next section. On the other hand, the central values of the LHCb and Belle 
results in \eqref{LHCb-CDF-BR} do not give real solutions for $x_s$. 
The requirement that the argument of the square-root in~(\ref{xs-BR-det}) 
is positive can be converted into 
the following lower bound:
\begin{equation}
\frac{ {\rm BR}(B_s\to D_s^{(*)\pm} K^\mp)_{\rm exp} } 
     { {\rm BR}(B_s\to D_s^{(*)\pm} \pi^\mp)_{\rm exp}} \geq 
     \frac{\epsilon}{ {\cal C}^{(*)}}
     \biggl[ 1 - y_s^2\, \cos^2\delta_s\,\cos^2(\phi_s+\gamma)\biggr] = 0.080 \pm 0.007,
     \label{BR-bound}
\end{equation}
which is shown in Fig.~\ref{fig:BRratio}.
We observe that the LHCb result for the ratio of branching ratios would need to increase by about
two standard deviations to satisfy this bound and to give a real solution for $x_s$.

In the next section, we shall use data from the  $B$ factories to obtain a
sharper picture of the hadronic parameters, including the CP-conserving strong phases
$\delta_s$.

\boldmath
\section{Hadronic Parameters from $B_d \to D_{(s)}^{(*)\pm} \pi^\mp$ Data}\label{sec:Bd}
\unboldmath
Using the $U$-spin flavour symmetry of strong
interactions, the hadronic parameters $x_s$ and $\delta_s$ of the $B_s\to D_s^{(*)\pm} K^\mp$ 
channels can be related to their counterparts $x_d$ and $\delta_d$ of the 
$B_d\to D^{(*)\pm} \pi^\mp$ decays as follows~\cite{RF-BsDsK}:
\begin{equation}
x_s = -\frac{x_d}{\epsilon}, \quad \delta_s = \delta_d.
\label{eq:xsxd}
\end{equation}
These relations assume exact $U$-spin symmetry; the impact of possible corrections
will be addressed below. 

The BaBar \cite{BaBar-Dpi} and Belle \cite{Belle-Dpi} collaborations have performed 
measurements which allow us to constrain the hadronic parameters $|x_d|$ and $\delta_s$.
For the $B_d\to D^\pm \pi^\mp$ system the following constraints have been extracted from
studies of CP-violating effects~\cite{HFAG}:
\begin{eqnarray}
a^{D\pi}           &\equiv & -2|x_d| \sin(\phi_d + \gamma) \cos(\delta_d) = 
-0.03 \pm 0.017,  \label{DpiCPV_a}\\
c_{\rm lep}^{D\pi} &\equiv & -2|x_d| \cos(\phi_d + \gamma) \sin(\delta_d) = 
-0.022 \pm 0.021. \label{DpiCPV}
\end{eqnarray}
A corresponding analysis of the $B_d\to D^{*\pm} \pi^\mp$ decays 
(for which $L=1$) yields \cite{HFAG}
\begin{align}
a^{D^*\pi}           &\equiv 2|x^V_d| \sin(\phi_d + \gamma) \cos(\delta^V_d) 
= -0.039 \pm 0.010, \label{DVpiCPV_a}\\
c_{\rm lep}^{D^*\pi} &\equiv 2|x^V_d| \cos(\phi_d + \gamma) \sin(\delta^V_d) 
= -0.010 \pm 0.013, \label{DVpiCPV}
\end{align}
where we have used the label $V$ to distinguish the vector $D^*$ system.
In order to convert these experimental results into $|x_d|$ and $\delta_d$, 
we assume  the value for $\gamma$ in (\ref{gam-det}) with the $B^0_d$--$\bar B^0_d$ 
mixing phase $\phi_d \equiv 2\beta = (42.8\pm 1.6)^\circ$ \cite{HFAG}, which yields 
$\phi_d + \gamma=(111\pm7)^\circ$.

Let us first extract $|x_d|$ by determining the doubly Cabibbo-suppressed branching ratio
${\rm BR}(\bar B_d^0 \to D^- \pi^+)$ from ${\rm BR}(\bar B_d^0\to D_s^-\pi^+)$
with the help of the $SU(3)$ flavour symmetry~\cite{dunietz}. Using the notation of
Ref.~\cite{fact-test}, we write
\begin{equation}\label{SU3-dun}
{\rm BR}(\bar B^0_d \to D^-\pi^+)=\left(\frac{\epsilon}{{\cal C}'}\right)
{\rm BR}(\bar B^0_d \to D_s^-\pi^+),
\end{equation}
where
\begin{equation}
{\cal C}'\equiv \frac{\Phi_{D_s\pi}}{\Phi_{D\pi}} {\cal N}'_F{\cal N}'_a{\cal N}'_E.
\end{equation}
In analogy to  (\ref{C-def}), the $\Phi$ are are phase-space factors, while 
\begin{equation}
 {\cal N}'_F \equiv \left[\frac{f_{D_s}}{f_D} \,
	                  \frac{F_1^{\bar B^0_d\pi^+}(m^2_{D_s})}{F_1^{\bar B^0_d\pi^+}(m^2_D)}\right]^2
\end{equation}
and
\begin{equation}
{\cal N}'_a \equiv  \left|\frac{a_1(D_s^+ \pi^-)}{a_1(D^+ \pi^-)} \right|^2
\end{equation}
describe factorizable and non-factorizable $SU(3)$-breaking effects, respectively. The 
${\cal N}'_E$ factor takes into account that $\bar B^0_d\to D^-\pi^+$ has a contribution
from an exchange topology, which does not have a counterpart in the $\bar B^0_d\to D_s^-\pi^+$
channel:
\begin{equation}
{\cal N}'_E \equiv  \left|\frac{T_{D^-\pi^+}}{T_{D^-\pi^+}+E_{D^-\pi^+}} \right|^2.
\end{equation}
We then obtain the following additional constraint for $x_d$:
\begin{equation} \label{eq:xdsquared}
|x_d| = \sqrt{\left(\frac{\epsilon}{{\cal C}'}\right)
\left[\frac{{\rm BR}(\bar B^0_d \to D_s^-\pi^+)}{{\rm BR}(\bar B^0_d \to D^+\pi^-)}\right]}.
\end{equation}

For the numerical analysis, we use the ratio of decay constants 
$f_{D_s}/f_{D} = 1.25\pm 0.06$~\cite{PDG} and
the form-factor ratio
$F_1^{\bar B^0_d\pi^+}(m^2_D)/F_1^{\bar B^0_d\pi^+}(m^2_{D_s}) = 0.9771 \pm 0.0009$,
where we have applied the evolution equation for the $\bar B^0_d\to\pi^+$ form factor given 
in Ref.~\cite{Duplancic:2008ix}. For the decays entering
(\ref{SU3-dun}), factorization is not expected to work well. Indeed, following the
approach discussed in Ref.~\cite{fact-test}, we extract 
$|a_1(D_s^+ \pi^-)| = 0.68\pm 0.12$ from the experimental data, while factorization
would correspond to a value around one. Unfortunately, an analogous factorization test for 
$\bar B_d^0 \to D^- \pi^+$ cannot be performed\footnote{The branching ratio quoted by the 
Particle Data Group \cite{PDG} is constructed from \eqref{eq:xdsquared}, 
so using this would create a circular 
argument.}. We allow for 20\% $SU(3)$-breaking effects for the non-factorizable
contributions, i.e.\ for the deviation of $|a_1|$ from one, leading to
${\cal N}'_a=1.0 \pm 0.2$.

In order to estimate the importance of the exchange contribution, we apply the
$SU(3)$ flavour symmetry and use
experimental information on $\mbox{BR}(\bar B^0_d\to D_s^+K^-)=(2.2\pm0.5)\times 10^{-5}$
\cite{PDG}, which receives only contributions from exchange topologies. 
Comparing it to the contribution from tree topologies, which we fix again through
${\rm BR}(\bar B_d^0 \to D_s^- \pi^+) = \left(2.16\pm 0.26\right)\times 10^{-5}$~\cite{B0Dspi},
we obtain:
\begin{equation}
\left|\frac{E_{D^-\pi^+}}{T_{D^-\pi^+}}\right|\sim 
\frac{f_\pi}{f_K} 
\left|\frac{V_{ub}}{V_{cb}}\right|
\sqrt{\frac{\mbox{BR}(\bar B^0_d\to D_s^+K^-)}{\mbox{BR}(\bar B^0_d \to D_s^-\pi^+)}}
\sim0.1.
\end{equation}
Consequently, we estimate ${\cal N}'_E \sim 1.0\pm0.2$. In comparison with the value of
${\cal N}_E \sim 0.97\pm0.08$ given after (\ref{NE-def}), this range is larger. Although
the exchange topologies entering both quantities are estimated to have similar absolute size, the
analysis performed in Ref.~\cite{fact-test} indicates a large angle between the $E$ and
$T$ amplitudes, which reduces the impact of $E$ on the amplitude ratio in ${\cal N}_E$.

Using finally also the experimental branching ratio
${\rm BR}(\bar B_d^0 \to D^+ \pi^-)   = \left(2.68\pm 0.13\right)\times 10^{-3}$~\cite{PDG}, 
the relation in \eqref{eq:xdsquared}  gives
\begin{equation}
	|x_d|=0.0163 \pm \left.0.0011\right|_{\rm BR} \left.\pm 0.0026\right|_{SU(3)} = 
	0.0163\pm 0.0028.	\label{DpiBR}
\end{equation}
This value is consistent with the results for $x_d$ given in Ref.~\cite{B0Dspi}. Combining 
(\ref{DpiBR}) with (\ref{DpiCPV_a}) and (\ref{DpiCPV}) allows, in principle, 
the determination of $\phi_d+\gamma$ and $\delta_d$ up to discrete ambiguities.
Unfortunately, a corresponding numerical fit leaves these parameters still largely 
unconstrained.

\begin{figure}[tbp] 
   \centering
   \includegraphics[width=10truecm]{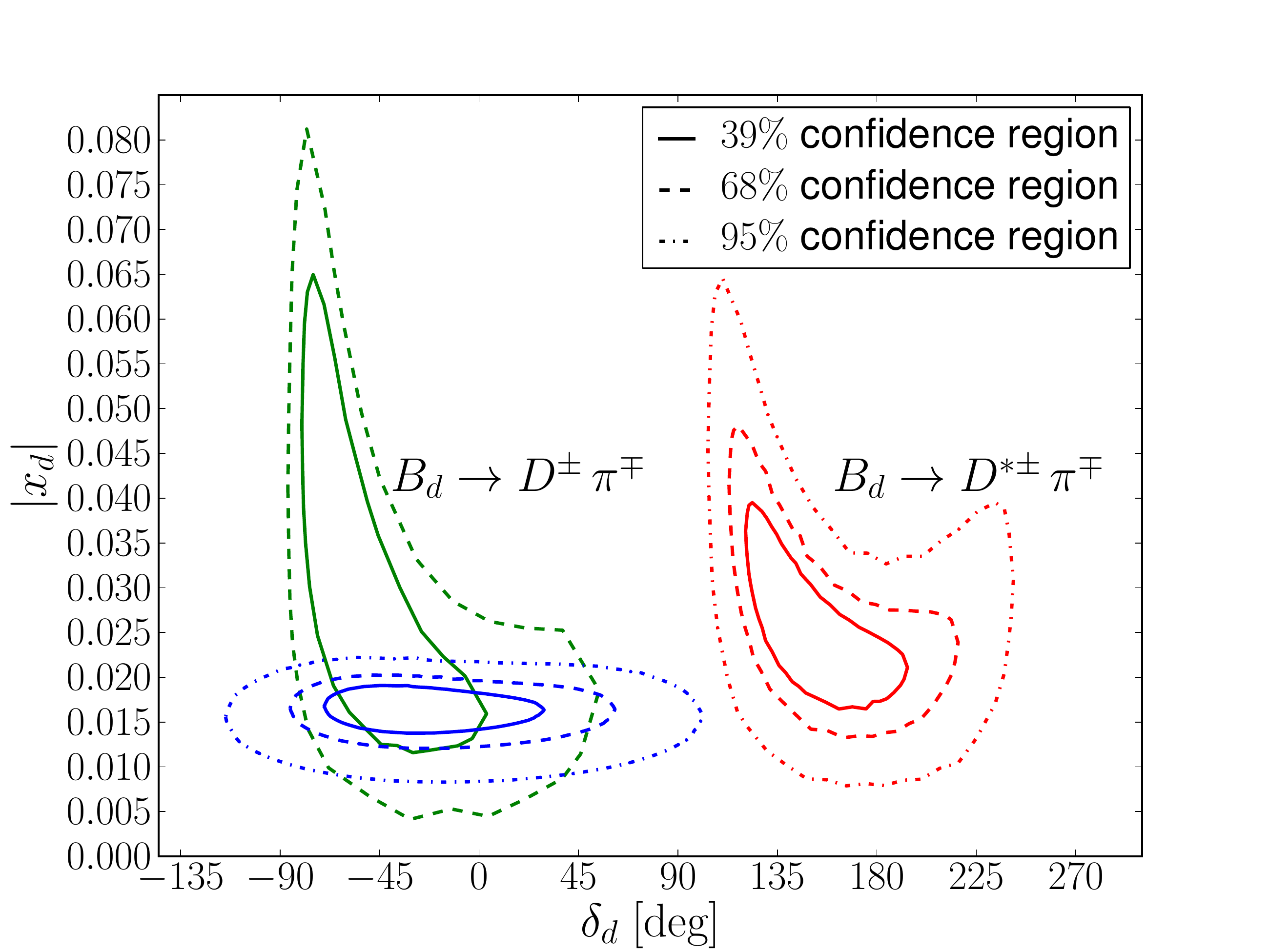} 
   \caption{The confidence level contours for the $\chi^2$ fit of the 
   hadronic parameters $|x_d|$ and $\delta_d$ as discussed in the text, illustrating also
   the impact of the $|x_d|$ constraint in (\ref{DpiBR}).}\label{fig:fit}
\end{figure}
We proceed to extract the parameters $|x_d|,\delta_d$ and $|x_d^V|,\delta_d^V$  from the 
constraints in \eqref{DpiCPV_a}--\eqref{DVpiCPV}  using a $\chi^2$ fit.
For the former parameter set we also include the constraint in \eqref{DpiBR}.
The fit gives the following results: 
\begin{align}
|x_d| &= 0.0166^{+0.0025}_{-0.0029}, & \delta_d &= \left(-35^{+65}_{-35}\right)^\circ, \\
|x^V_d| &= 0.025^{+0.014}_{-0.008}, & \delta^V_d &= \left(146^{+48}_{-25}\right)^\circ,
\end{align}
where the errors give the $68\%$ confidence level for each parameter.
The $\chi^2/n_{\rm dof}$ is 0.53 and 0.00 for the non-vector and vector decays, respectively.
In Fig.~\ref{fig:fit}, we show the corresponding  39\%, 68\% and 95\% confidence level regions 
in the $\delta_d$--$|x_d|$ plane.
Note that the constraint in \eqref{DpiBR} considerably reduces the uncertainty of the 
$|x_d|$ parameter for the non-vector decay.

Using (\ref{eq:xsxd}), we hereby find
\begin{align}\label{uspinRes}
	x_s &= 0.311^{+0.046}_{-0.053}\big|_{\rm input}\pm 0.06\big|_{SU(3)}, & 	
	\delta_s &= \left[-35^{+69}_{-40}\big|_{\rm input}\pm 20 \big|_{SU(3)}\right]^\circ,\\
	x^V_s &= 0.47^{+0.26}_{-0.15}\big|_{\rm input}\pm 0.09\big|_{SU(3)}, & 
	\delta^V_s &= \left[146^{+48}_{-25}\big|_{\rm input}\pm 20 \big|_{SU(3)}\right]^\circ,
\label{uspinResV}
\end{align}
where we allow for $SU(3)$-breaking effects of 20\% for the 
$x^{(V)}_s$ parameters and $\pm 20^\circ$ for the strong phases.
In later applications of these results,  the uncertainties associated with the 
$x^{(V)}_d$, $\delta^{(V)}_d$ parameters and the $SU(3)$-breaking effects
will be combined in quadrature.

Before using the hadronic parameters given above to predict the observables of the 
$B_s\to D_s^{(*)\pm} K^\mp$ decays in Section~\ref{sec:exp}, which serve as input
for an experimental study, let us first discuss the extraction of $\phi_s+\gamma$
from these channels, with a special emphasis on multiple discrete ambiguities and their
resolution.

\boldmath
\section{Extraction of $\phi_s+\gamma$ and Discrete Ambiguities}\label{sec:CP}
\unboldmath
For the extraction of $\phi_s+\gamma$ from the $B_s\to D_s^{(*)\pm} K^\mp$ system,
it is necessary to measure the following CP asymmetries from time-dependent, tagged 
analyses:
\begin{align}
	a_{\rm CP}(B_s(t)\to D_s^{(*)+} K^-) 
	&\equiv \frac{\Gamma(B^0_s(t)\to D_s^{(*)+} K^-) - \Gamma(\bar{B}^0_s(t)\to D_s^{(*)+} K^-) }
	{\Gamma(B^0_s(t)\to D_s^{(*)+} K^-) + \Gamma(\bar{B}^0_s(t)\to D_s^{(*)+} K^-) } \notag\\
	&= \frac{{C}\,\cos(\Delta M_s\,t) + {S}\,\sin(\Delta M_s\,t)}
	{\cosh(y_s\,t/\tau_{B_s}) + {\cal A}_{\Delta\Gamma}\,\sinh(y_s\,t/\tau_{B_s})};
	\label{ACP}
\end{align}
an analogous expression holds for the CP-conjugate $D_s^{(*)-} K^+$ final states, where $C$, 
$S$ and ${\cal A}_{\Delta\Gamma}$ are simply replaced by $\overline{C}$, $ \overline{S}$
and $\overline{{\cal A}}_{\Delta\Gamma}$, respectively. 
The observables take the following form \cite{RF-BsDsK}:
\begin{equation} \label{Casym-def}
	C = -\left[\frac{1-x_s^2}{1+x_s^2}\right],   \quad  \overline{C} 
	= +\left[\frac{1-x_s^2}{1+x_s^2}\right] 
\end{equation}	
\begin{equation} 
	S = (-1)^L \frac{2\,x_s}{1+x_s^2} \sin(\phi_s + \gamma + \delta_s), \quad
	\overline{S} = (-1)^L \frac{2\,x_s}{1+x_s^2} \sin(\phi_s + \gamma - \delta_s),
\end{equation}	
which complement the expressions for ${\cal A}_{\Delta\Gamma}$ and 
$\overline{{\cal A}}_{\Delta\Gamma}$ in (\ref{ADG-expr}).

For the following discussion, it is convenient to introduce the observable combinations 
\begin{equation}
\langle C \rangle_\pm \equiv \frac{\overline{C} \pm C}{2}, \quad
\langle S \rangle_\pm \equiv \frac{\overline{S} \pm S}{2},
\label{eq:CSAcomb}
\end{equation}
as well as 
\begin{equation}\label{splus}
s_+\equiv(-1)^L\left[\frac{1+x_s^2}{2\,x_s}\right]\langle S\rangle_+
=+\cos\delta_s\sin(\phi_s+\gamma)
\end{equation} 
\begin{equation}\label{s-m}
s_-\equiv(-1)^L\left[\frac{1+x_s^2}{2\,x_s}\right]\langle S\rangle_-
=-\sin\delta_s\cos(\phi_s+\gamma),
\end{equation} 
where 
\begin{equation}\label{xs-def}
x_s=\sqrt{\frac{1-\langle C\rangle_-}{1+\langle C\rangle_-}}, \quad\mbox{yielding}\quad
\frac{1+x_s^2}{2\,x_s}=\frac{1}{\sqrt{1-\langle C\rangle_-^2}}.
\end{equation}
Finally, we obtain
\begin{equation}\label{conv}
\sin(\phi_s+\gamma)=\pm\sqrt{\frac{1}{2}\left[(1+s_+^2-s_-^2)\pm
\sqrt{(1+s_+^2-s_-^2)^2-4s_+^2}\right]},
\end{equation}
which results in an eightfold solution for $\phi_s+\gamma$ \cite{BsDsK,RF-BsDsK}. 

As was pointed out in Ref.~\cite{RF-BsDsK}, and later also in 
Refs.~\cite{Cavoto:2006um,Gligorov:2008zzb,Gligorov:2011id}, the observable combinations
\begin{equation}
\langle {\cal A}_{\Delta\Gamma} \rangle_+ \equiv
	\frac{\overline{\cal A}_{\Delta\Gamma} + {\cal A}_{\Delta\Gamma}}{2}
= -(-1)^L\left[\frac{2\,x_s}{1+x_s^2}\right] \cos\delta_s\cos(\phi_s+\gamma)
\end{equation}
\begin{equation}
\langle {\cal A}_{\Delta\Gamma} \rangle_- \equiv
	\frac{\overline{\cal A}_{\Delta\Gamma} - {\cal A}_{\Delta\Gamma}}{2}
= -(-1)^L\left[\frac{2\,x_s}{1+x_s^2}\right] \sin\delta_s\sin(\phi_s+\gamma)
\end{equation}
can be combined with the mixing-induced CP asymmetries $\langle S \rangle_\pm$ 
to derive the relation
\begin{equation}
	\tan(\phi_{s} + \gamma) =
	-\frac{\langle S \rangle_+}{\langle {\cal A}_{\Delta\Gamma} \rangle_+} 
	= \frac{\langle {\cal A}_{\Delta\Gamma} \rangle_-}{\langle S \rangle_-},
	\label{extractPhiW}
\end{equation}
which allows the extraction of $\phi_s+\gamma$ up to a twofold ambiguity; 
moreover, we have
\begin{equation}\label{tan-rel-2}
|\tan(\phi_{s} + \gamma)|=\sqrt{\frac{\langle {\cal A}_{\Delta\Gamma} \rangle^2_-+
\langle S \rangle^2_+}{\langle S \rangle^2_- + {\langle {\cal A}_{\Delta\Gamma} \rangle^2_+}}}=
\sqrt{\frac{\langle {\cal A}_{\Delta\Gamma} \rangle^2_- -
\langle S \rangle^2_+}{\langle S \rangle^2_- -{\langle {\cal A}_{\Delta\Gamma} \rangle^2_+}}}.
\end{equation}
The final ambiguity can be resolved from factorization arguments, where we expect
\begin{equation}
	\cos\delta_s > 0,\quad \cos\delta_s^V < 0,
\end{equation}
a pattern that agrees well with the results of the $U$-spin analysis presented 
in Section~\ref{sec:Bd}, where
the results for the strong phases in (\ref{uspinRes}) and (\ref{uspinResV}) give
\begin{equation}
\cos\delta_s=  0.82^{+0.18}_{-0.56} , \quad \cos\delta_s^V= -0.83^{+0.43}_{-0.17}  .
\end{equation}
Combining this with (\ref{splus}), the sign of $\sin(\phi_s+\gamma)$ can then 
be determined. Thus, under reasonable assumptions, the extraction of 
$\phi_{s} + \gamma$ is unambiguous \cite{RF-BsDsK}.  
A discussion of the resolution of these discrete ambiguities was also 
given in Ref.~\cite{NandiNierste_DsK}.

\begin{figure}[tbp] 
   \centering
   \begin{tabular}{cc}
	   \includegraphics[width=7.4truecm]{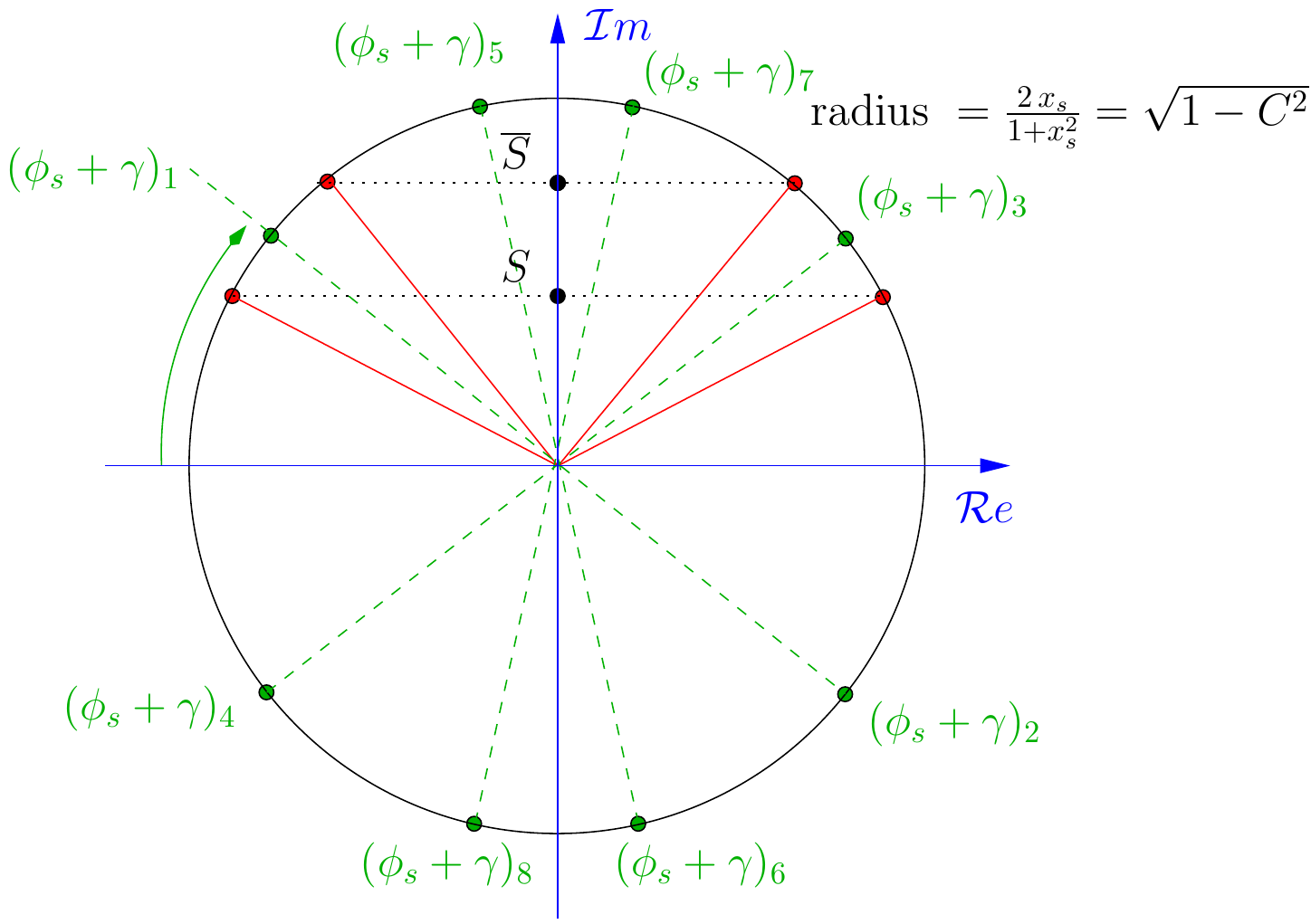} &
	   \includegraphics[width=7.4truecm]{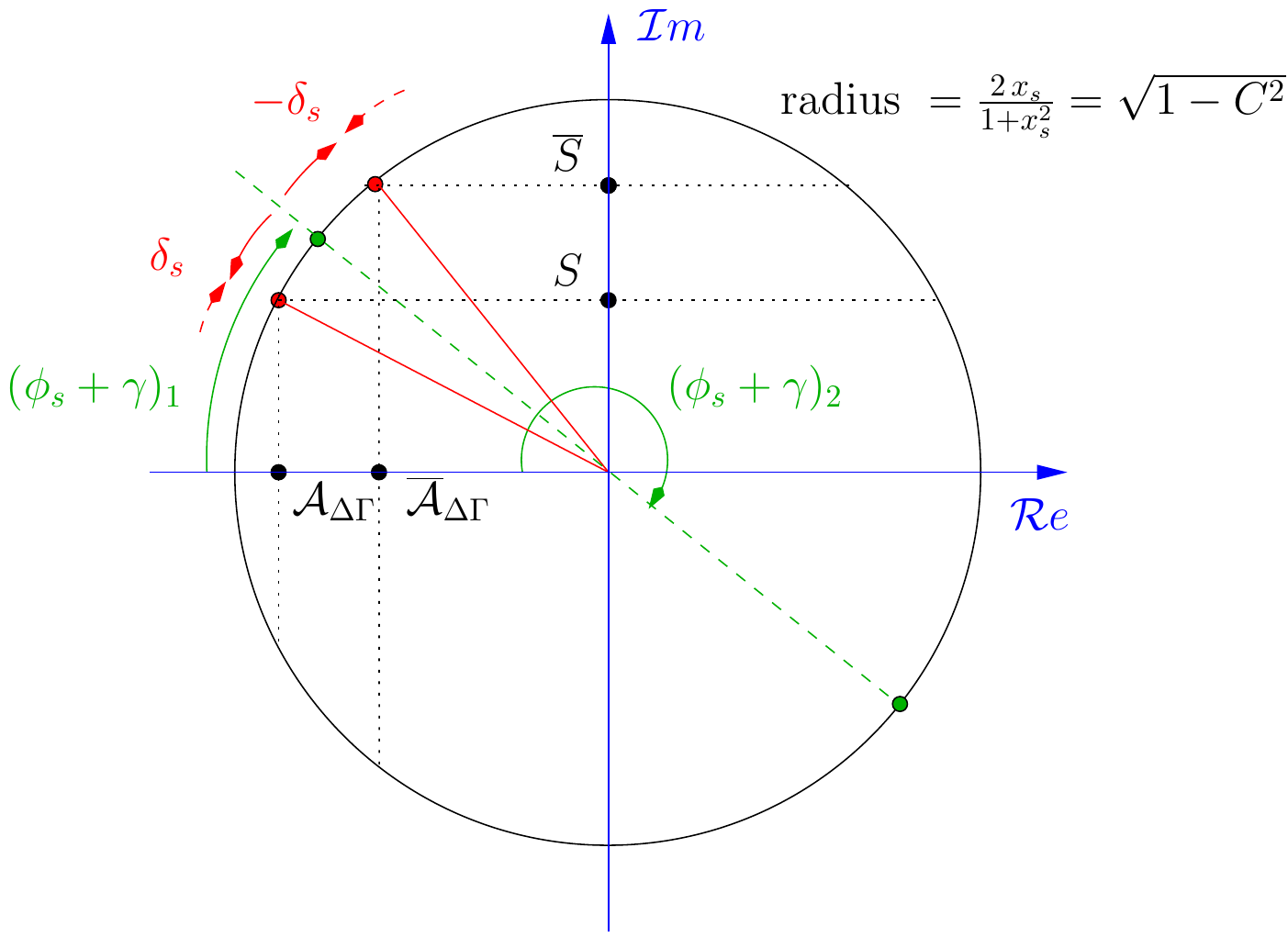} 
    \end{tabular}
   \caption{Illustration of  the complex numbers $({\cal A}_{\Delta\Gamma}+iS)$ and 
   $(\overline{\cal A}_{\Delta\Gamma} + i\,\overline{S})$  with lengths $\sqrt{1-C^2}$
   in the complex plane.
   Left panel: illustration of the extraction of $\phi_s+\gamma$ without 
   the use of untagged information and the
   associated eightfold discrete ambiguity (see (\ref{conv})). Right panel: illustration of the
   reduction of the discrete ambiguity to a twofold one through the untagged observables 
   ${\cal A}_{\Delta\Gamma}$ and  $\overline{\cal A}_{\Delta\Gamma}$ (see (\ref{extractPhiW})).}
   \label{fig:polar}
\end{figure}

It is instructive to illustrate these features, which can be hidden in a global 
experimental fit (see Section~\ref{sec:exp}). As the observables satisfy 
\begin{align}
	{C}^2 + S^2 + {\cal A}^2_{\Delta\Gamma} = 1
	= \overline{C}^2 + \overline{S}^2 + \overline{\cal A}_{\Delta\Gamma}^2 
	\label{dependence}
\end{align}
and are hence not independent, only two of the three observables for each of the final states 
of the $B_s \to D_s^{(*)\pm} K^\mp$ system are needed for the determination of $\phi_s + \gamma$. 
We introduce the complex 
numbers\footnote{Their relation to the complex observables $\xi$ and $\overline{\xi}$ 
defined in Ref.~\cite{RF-BsDsK} is given by $2\,{\xi}/(1+|{\xi}|^2) = 
{\cal A}_{\Delta\Gamma} + i\,{S}$ and $2\,{\overline{\xi}}/(1+|\overline{\xi}|^2) = 
\overline{\cal A}_{\Delta\Gamma} + i\,\overline{S}$, respectively.} 
\begin{align}
	{\cal A}_{\Delta\Gamma} + i\,{S} &= -(-1)^L \sqrt{1 - C^2}\, 
	e^{-i(\phi_{s} + \gamma + \delta_s)}\label{vec1} \\
	\overline{\cal A}_{\Delta\Gamma} + i\,\overline{S} &= 
	-(-1)^L \sqrt{1 - \overline{C}^2}\, e^{-i(\phi_{s} + \gamma - \delta_s)} \label{vec2},
\end{align}
which, as $C = -\overline{C}$ (see (\ref{Casym-def})), have the same absolute value and 
thus span the same circle in the complex plane.  The weak phase $\phi_s + \gamma$ 
corresponds to the polar angle of a complex number that lies exactly between 
\eqref{vec1} and \eqref{vec2}, with an equal angular distance of $\delta_s$ to both.

Let us first have a look at the strategy, which does not use the information 
provided by the untagged ${\cal A}_{\Delta\Gamma}$,  $\overline{\cal A}_{\Delta\Gamma}$ 
observables. The $C$ and $\overline{C}$  then fix a circle in the complex plane, while the 
mixing-induced CP asymmetries $S$, $\overline{S}$ fix the component in the imaginary 
direction. As illustrated in the left panel of Fig.~\ref{fig:polar}, this results in an eightfold 
discrete ambiguity for $\phi_s + \gamma$. On the contrary, as shown in the right panel of 
Fig.~\ref{fig:polar}, if the mixing-induced CP asymmetries are measured together with the 
untagged observables, the discrete ambiguity is reduced to a twofold one, which can be 
fully resolved as discussed above. Consequently, the optimal observable sets for 
the extraction of $\phi_s+\gamma$ are 
$S$, $\overline{S}$ and ${\cal A}_{\Delta\Gamma}$,  $\overline{\cal A}_{\Delta\Gamma}$. 

Another important advantage of these observables is not only that they depend 
linearly on $x_s$ -- in contrast to $C$, $\overline{C}$ and the determination 
of this parameter through (\ref{xs-BR-det}) -- but that $x_s$ drops out
in (\ref{extractPhiW}) and (\ref{tan-rel-2}). Interestingly, as we will see in the next 
section, both observable sets can be accessed with similar precision at LHCb: 
the extraction of the untagged 
${\cal A}_{\Delta\Gamma}$,  $\overline{\cal A}_{\Delta\Gamma}$ observables relies 
on the $B_s$ decay width parameter (\ref{ys-LHCb}), while the measurement of the 
$S$, $\overline{S}$ observables requires the tagging of the flavour of the initially 
produced $B^0_s$ or $\bar{B}_s^0$ mesons.

\section{Experimental Prospects}\label{sec:exp}
The hadronic parameters determined in Section~\ref{sec:Bd}, with the phases 
in (\ref{phis-average}) and (\ref{gam-det}), allow us to make predictions
of the observables of the $B_s\to D_s^{\pm} K^\mp$ decays:
\begin{align}
	\tau_{\rm eff}          &= 0.971^{+0.053}_{-0.012}\ \tau_{B_s}, &
	{\cal A}_{\Delta\Gamma} &= -0.49^{+0.58}_{-0.13}, & 
        C                       &= -0.824^{+0.086}_{-0.077},& 
	S                       &= 0.29^{+0.30}_{-0.40},    \notag\\
	\bar\tau_{\rm eff}               &= 1.025^{+0.030}_{-0.054}\ \tau_{B_s}, & 
        \overline{\cal A}_{\Delta\Gamma} &= 0.11^{+0.34}_{-0.59}, & 
        \overline{C}                     &= 0.824^{+0.077}_{-0.086}, &  
        \overline{S}                     &= 0.55^{+0.11}_{-0.28}. 
	\label{eq:theo_pred}
\end{align}
In analogy, for the $B_s\to D_s^{*\pm} K^\mp$ decays we obtain
\begin{align}
	 \tau^V_{\rm eff}          &= 0.954^{+0.057}_{-0.021}\ \tau_{B_s}, &
	 {\cal A}_{\Delta\Gamma}^V &= -0.66^{+0.60}_{-0.21}, &  
	 C^V                       &= -0.64^{+0.36}_{-0.20}, & 
         S^V                       &= 0.40^{+0.39}_{-0.44},    \notag\\
	\bar\tau^V_{\rm eff}               &= 1.027^{+0.034}_{-0.060}\ \tau_{B_s}, &
	\overline{\cal A}_{\Delta\Gamma}^V &= 0.13^{+0.40}_{-0.66}, & 
	\overline{C}^V                     &= 0.64^{+0.20}_{-0.36}, & 
	\overline{S}^V                     &= 0.76^{+0.19}_{-0.30}.  
\end{align}
Furthermore, our predictions for the branching ratio observables \eqref{LHCb-CDF-BR}
and \eqref{BR-diff} are
\begin{align}
	\left.\frac{\text{BR}(B_s\rightarrow 
	D_s^\pm K^\mp)_{\rm exp}}{\text{BR}(B_s\rightarrow 
	D_s^\pm \pi^\mp)_{\rm exp}}\right|_{SU(3)}
	&= 0.0864^{+0.0087}_{-0.0072}, \label{eq:theo_predBR}\\
	\left.\frac{{\rm BR}(B_s\to D_s^{+} K^-)_{\rm exp}-
{\rm BR}(B_s\to D_s^{-} K^+)_{\rm exp}}{{\rm BR}(B_s\to D_s^{+} K^-)_{\rm exp}+
{\rm BR}(B_s\to D_s^{-} K^+)_{\rm exp}}\right|_{SU(3)} &= -0.027^{+0.052}_{-0.019},
\end{align}
respectively. The prediction in  (\ref{eq:theo_predBR}) is compared to the current experimental
results in Fig.~\ref{fig:BRratio}. Similarly, we predict for the vector decays:
\begin{align}
	\left.\frac{\text{BR}(B_s\rightarrow 
	D_s^{*\pm} K^\mp)_{\rm exp}}{\text{BR}(B_s\rightarrow 
	D_s^{*\pm} \pi^\mp)_{\rm exp}}\right|_{SU(3)}
	&= 0.099^{+0.030}_{-0.036}, \\
	\left.\frac{{\rm BR}(B_s\to D_s^{*+} K^-)_{\rm exp}-
{\rm BR}(B_s\to D_s^{*-} K^+)_{\rm exp}}{{\rm BR}(B_s\to D_s^{*+} K^-)_{\rm exp}+
{\rm BR}(B_s\to D_s^{*-} K^+)_{\rm exp}}\right|_{SU(3)} &= -0.035^{+0.056}_{-0.024}.
\end{align}

To estimate the experimental sensitivity for the observables,
a simple Monte Carlo simulation has been performed, using 
as theoretical input the central values $x_s=0.311$, $\delta_s=-35^{\circ}$ (see (\ref{uspinRes})),
$\Delta m_s=17.72\,\mathrm{ps}^{-1}$~\cite{LHCb-CONF-2011-050-deltams}, 
$y_s=0.088$ (see \eqref{ys-LHCb}) and
$\gamma+\phi_s=65.5^{\circ}$ (see (\ref{phis-average}) and (\ref{gam-det})).
A global fit to the decay distributions then simultaneously determines the observables given in
(\ref{eq:theo_pred}).

\begin{table}[tb]
\caption{Statistical uncertainties of $B_s^0\rightarrow D_s^{\pm}K^{\mp}$ 
CP observables for various data samples as determined from our toy study. 
The difference in sensitivity of $\langle \mathcal{A}_{\Delta\Gamma}\rangle_+$, 
$\langle \mathcal{A}_{\Delta\Gamma}\rangle_-$ is due to a correlation 
between $\mathcal{A}_{\Delta\Gamma}$ and $\overline{\mathcal{A}}_{\Delta\Gamma}$ 
of 0.5 observed in our toy simulations.}
\label{Tab:toy}
\vspace*{-0.4truecm}
\center
\begin{tabular}{l|cccccc}
\toprule
Scenario & $\sigma(C\:,\:\overline{C})$ & $\sigma(S\:,\:\overline{S})$ 
& $\sigma(\mathcal{A}_{\Delta\Gamma}\:,\overline{\mathcal{A}}_{\Delta\Gamma})$ 
& $\sigma(\langle S\rangle_{\pm})$ & $\sigma(\langle \mathcal{A}_{\Delta\Gamma}\rangle_+)$ 
& $\sigma(\langle \mathcal{A}_{\Delta\Gamma}\rangle_-)$\\
\midrule
LHCb end 2012 & $\pm 0.176$ & $\pm 0.252$ 
& $\pm 0.210$ & $\pm 0.173$ & $\pm 0.194$ & $\pm 0.113$\\
LHCb 2018 & $\pm 0.077$ & $\pm 0.110$ & $\pm 0.092$ 
& $\pm 0.076$ & $\pm 0.085$ & $\pm 0.049$\\
LHCb Upgrade & $\pm 0.032$ & $\pm 0.046$ & $\pm 0.038$ &
 $\pm 0.032$ & $\pm 0.035$ & $\pm 0.020$\\
\bottomrule
\end{tabular}
\end{table}

\begin{table}[tb]
\caption{Experimental uncertainties on the weak phase $\phi_s + \gamma$, strong 
phase $\delta_s$ and hadronic parameter $x_s$ for various data samples as determined 
from our toy simulations. Results for the method, which excludes the untagged 
observables, are also shown. The errors correspond to the central values 
$\phi_s + \gamma=65.5^\circ$, $\delta_s=-35^\circ$ and $x_s=0.31$.}
\label{Tab:toy2}
\vspace*{-0.2truecm}
\center
\begin{tabular}{l|ccc|ccc}
\toprule
& \multicolumn{3}{c|}{With $\mathcal{A}_{\Delta\Gamma}$ 
and $\overline{\mathcal{A}}_{\Delta\Gamma}$} & \multicolumn{3}{c}{Only tagged information}\\
\midrule
Scenario  & $\phi_s + \gamma$ & $\delta_s$ & $x_s$  & $\phi_s + \gamma$ & $\delta_s$ & $x_s$\\
\midrule
LHCb end 2012 & $[\pm 17]^{\circ}$ & $[\pm 17]^{\circ}$ 
& $\pm 0.080$ & - & - & $\pm 0.11\phantom{0}$\\
LHCb 2018 & $[\pm 7.3]^{\circ}$ & $[\pm 7.3]^{\circ}$& 
$\pm 0.035$   & $[^{+16}_{-26}]^{\circ}$ & $[^{+26}_{-16}]^{\circ}$ & $\pm 0.048$\\
LHCb Upgrade  & $[\pm 3.0]^{\circ}$ & $[\pm 3.0]^{\circ}$ & 
$\pm 0.015$ & $[^{+8.8}_{-19}]^{\circ}$ & $[^{+19}_{-8.8}]^{\circ}$ & $\pm 0.021$ \\
\bottomrule
\end{tabular}
\end{table}

\begin{figure}
\center
\includegraphics[width=0.32\textwidth]{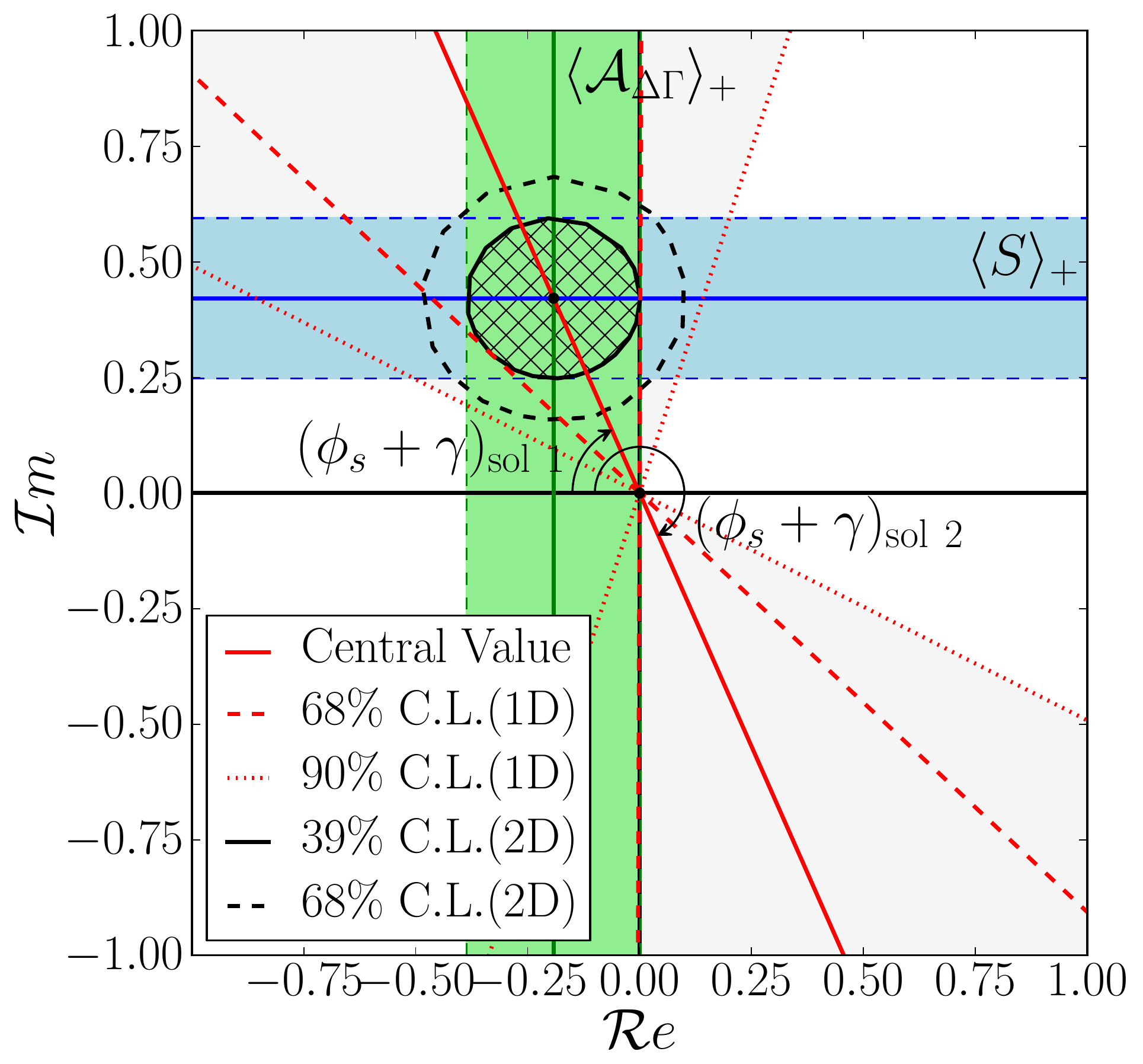}
\includegraphics[width=0.32\textwidth]{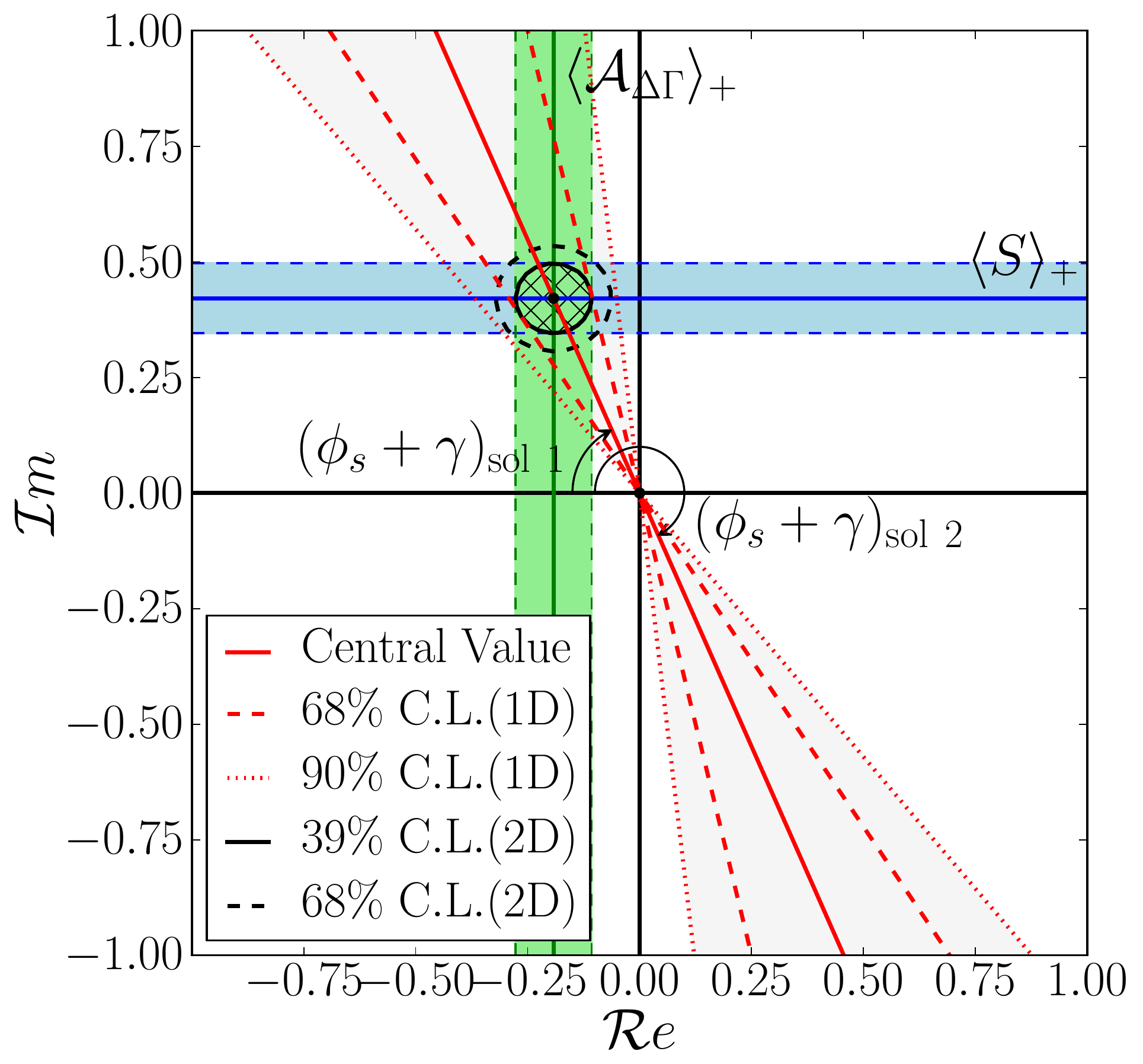}
\includegraphics[width=0.32\textwidth]{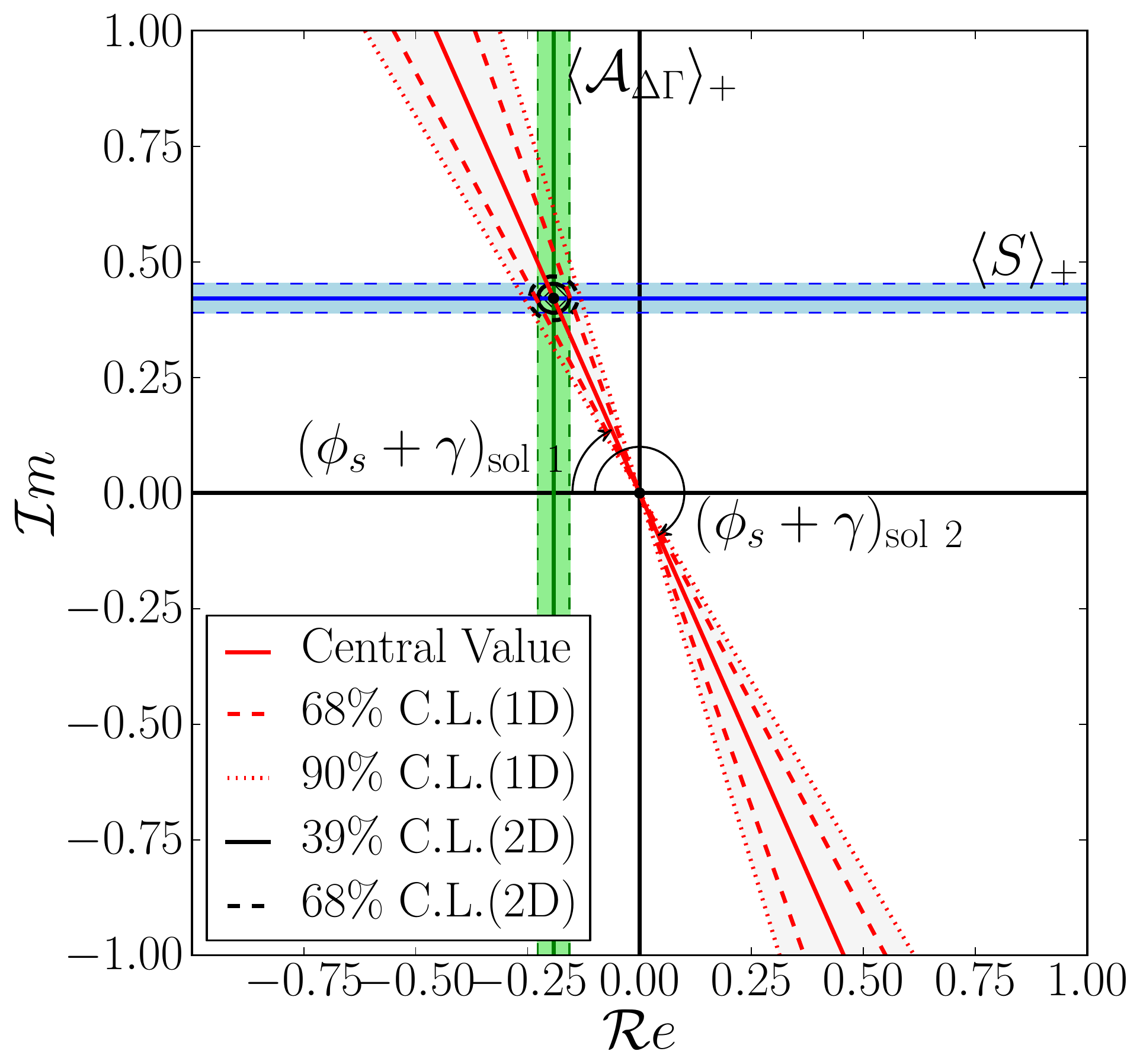}
\includegraphics[width=0.32\textwidth]{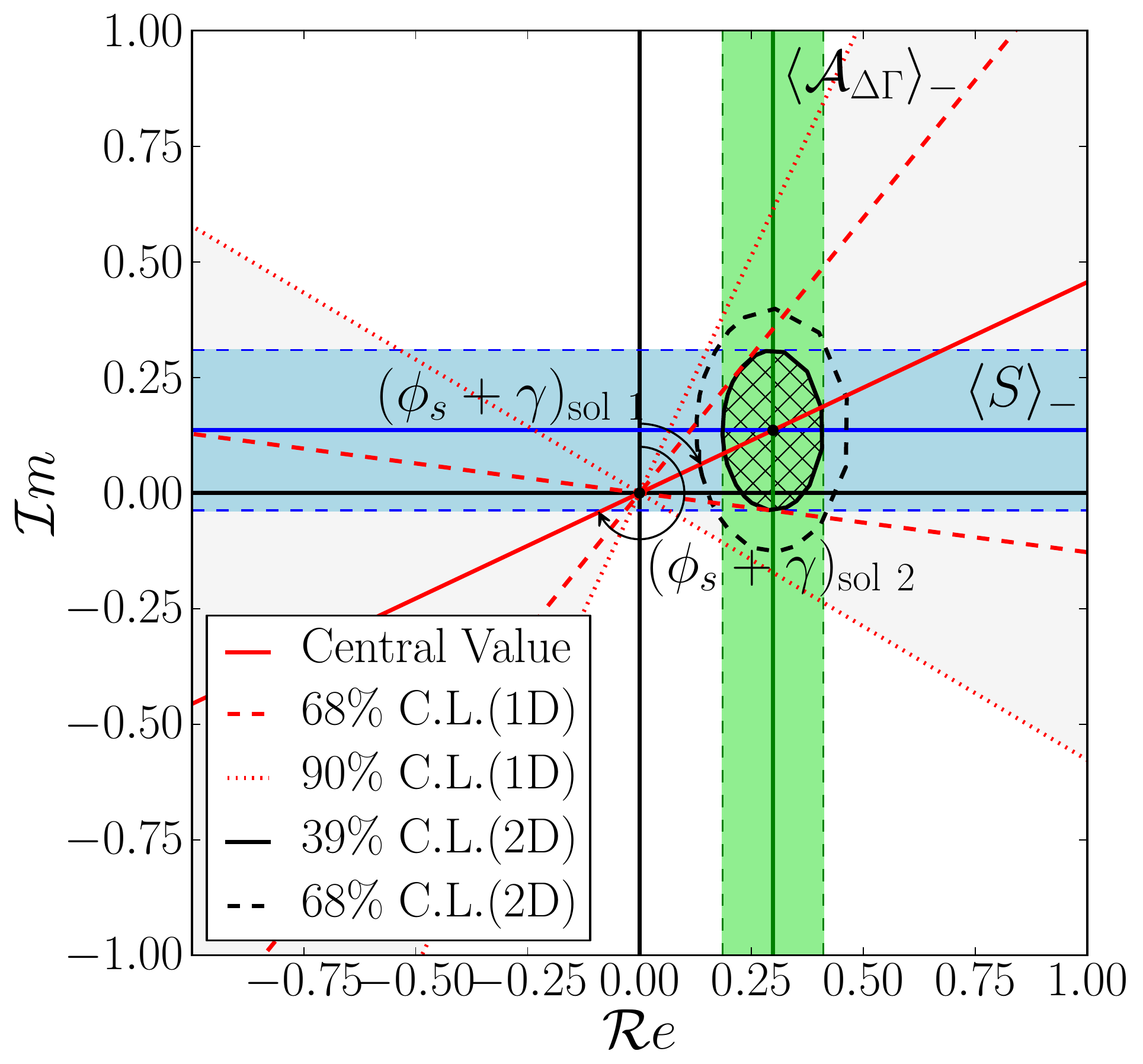}
\includegraphics[width=0.32\textwidth]{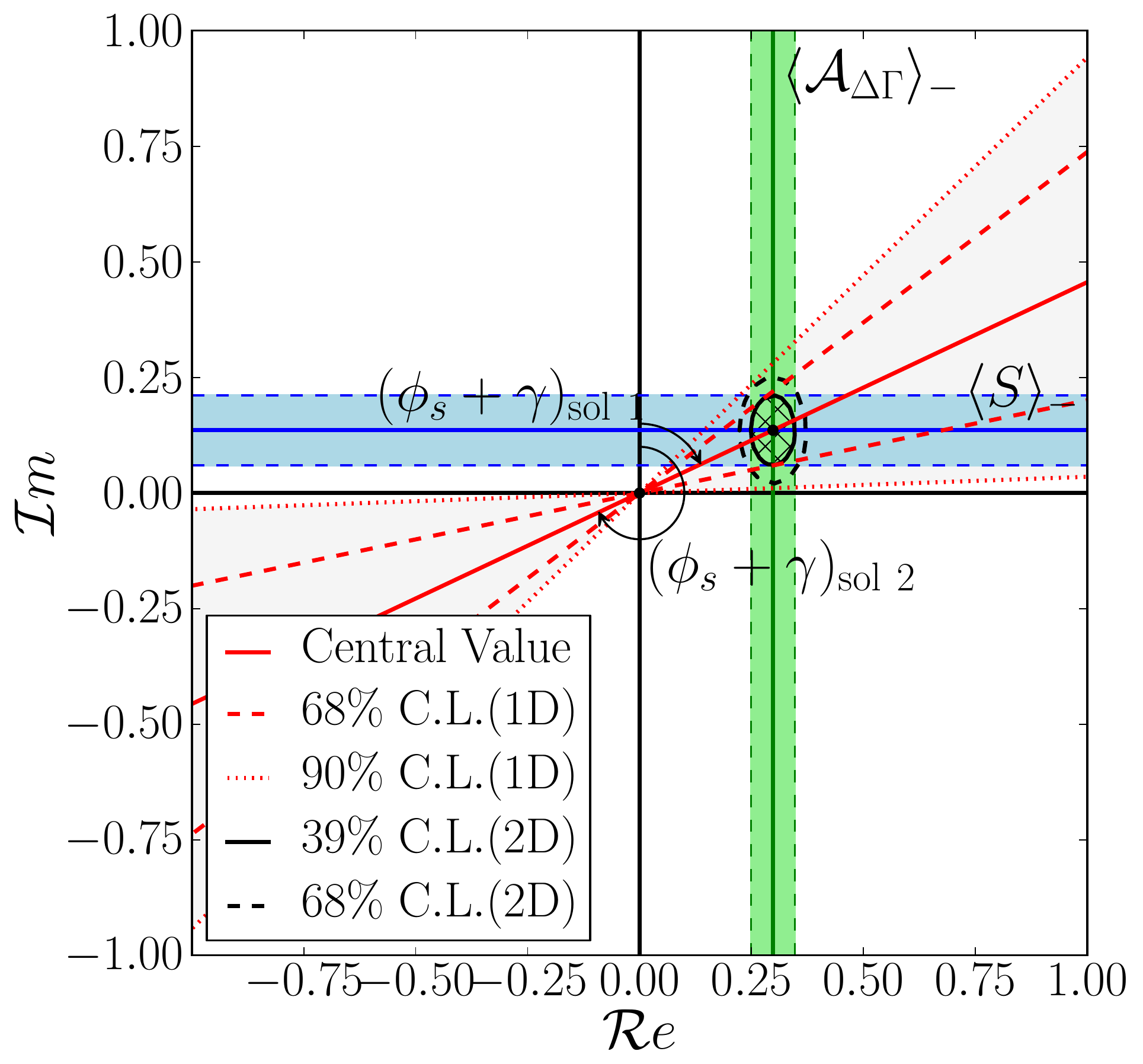}
\includegraphics[width=0.32\textwidth]{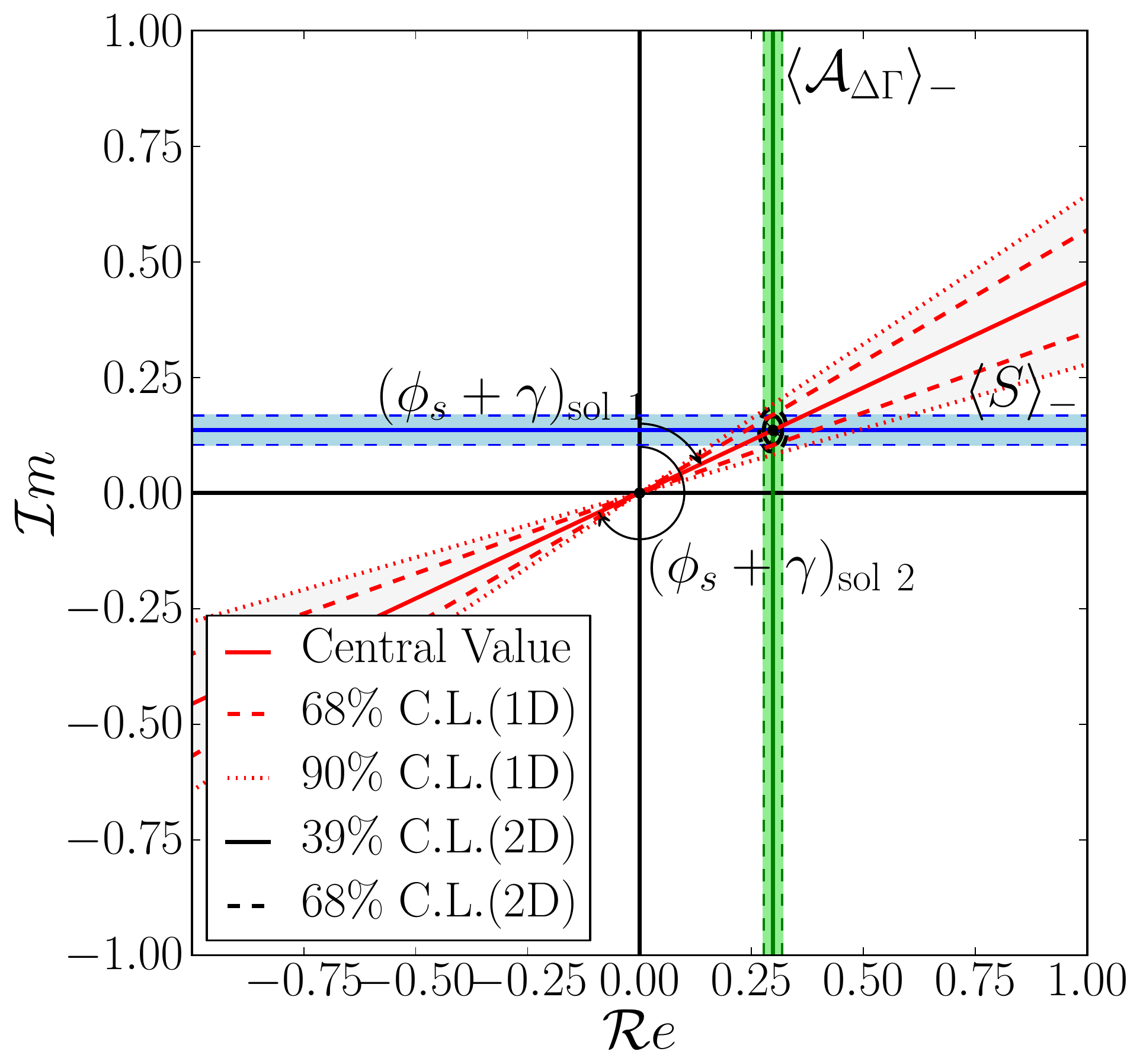}
\caption{Illustration of the determination of $\phi_s+\gamma$ from the separate observable 
combinations $\langle {\cal A}_{\Delta\Gamma}\rangle_+$ and $\langle S\rangle_+$ 
(top row) and $\langle {\cal A}_{\Delta\Gamma}\rangle_-$ and $\langle S\rangle_-$ 
(bottom row); see (\ref{comp-1}) and (\ref{comp-2}).
The increasing experimental sensitivity of the panels from left to right corresponds to 
expectations of the LHCb experiment by the end of 2012, before the upgrade and after the 
upgrade, respectively.}
\label{fig:toy_angle}
\end{figure}

\begin{figure}
\center
\includegraphics[width=0.32\textwidth]{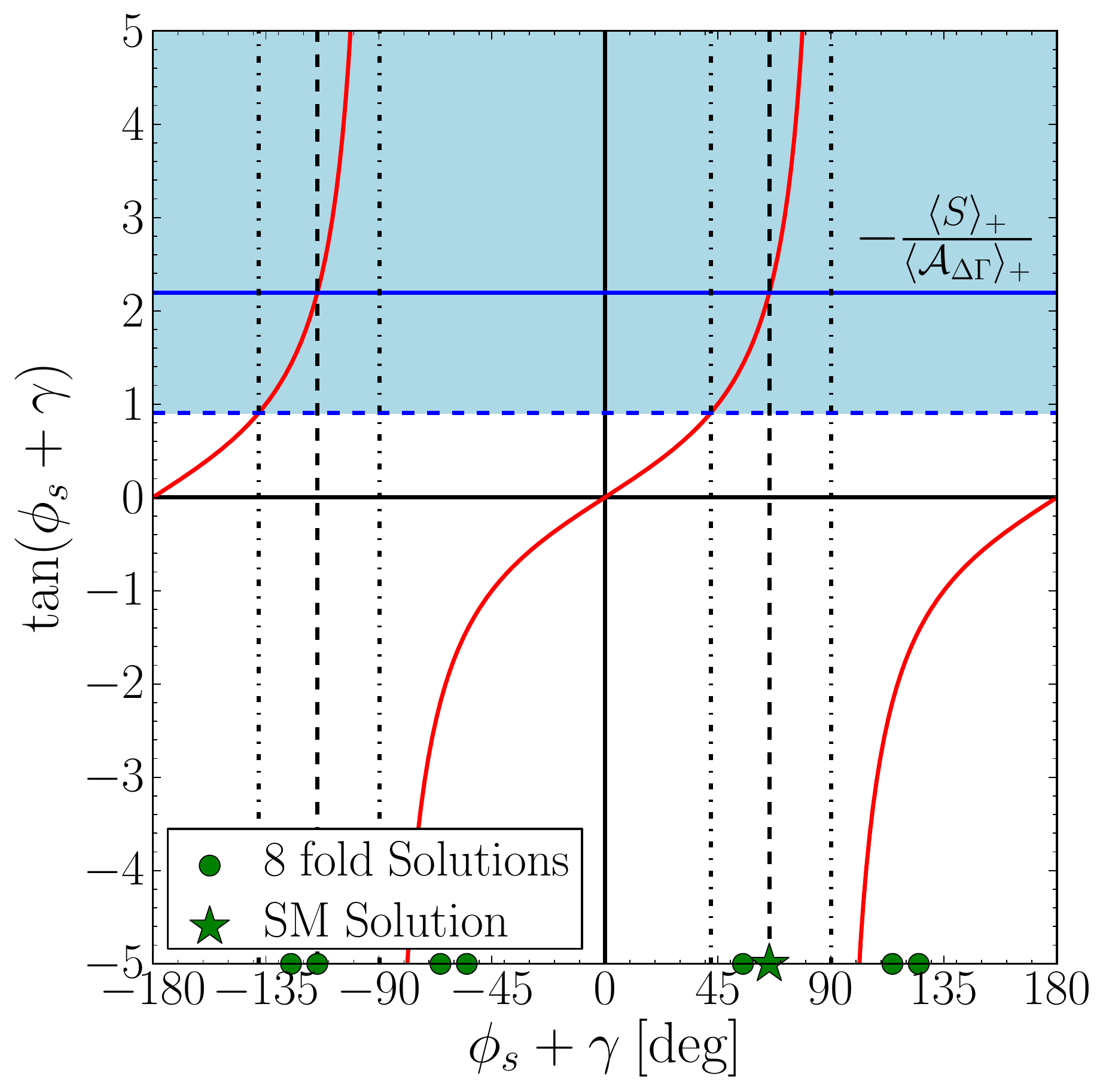}
\includegraphics[width=0.32\textwidth]{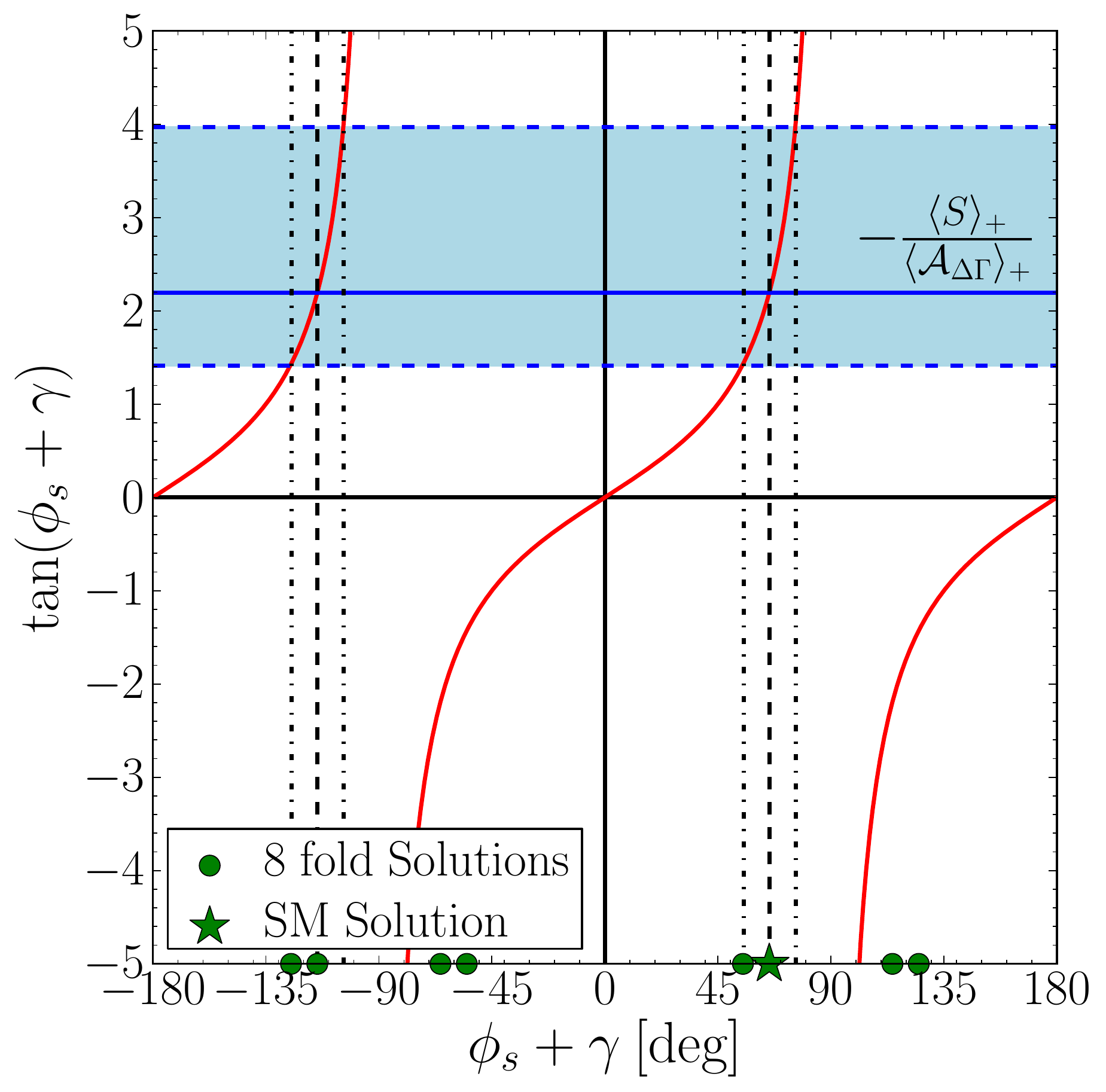}
\includegraphics[width=0.32\textwidth]{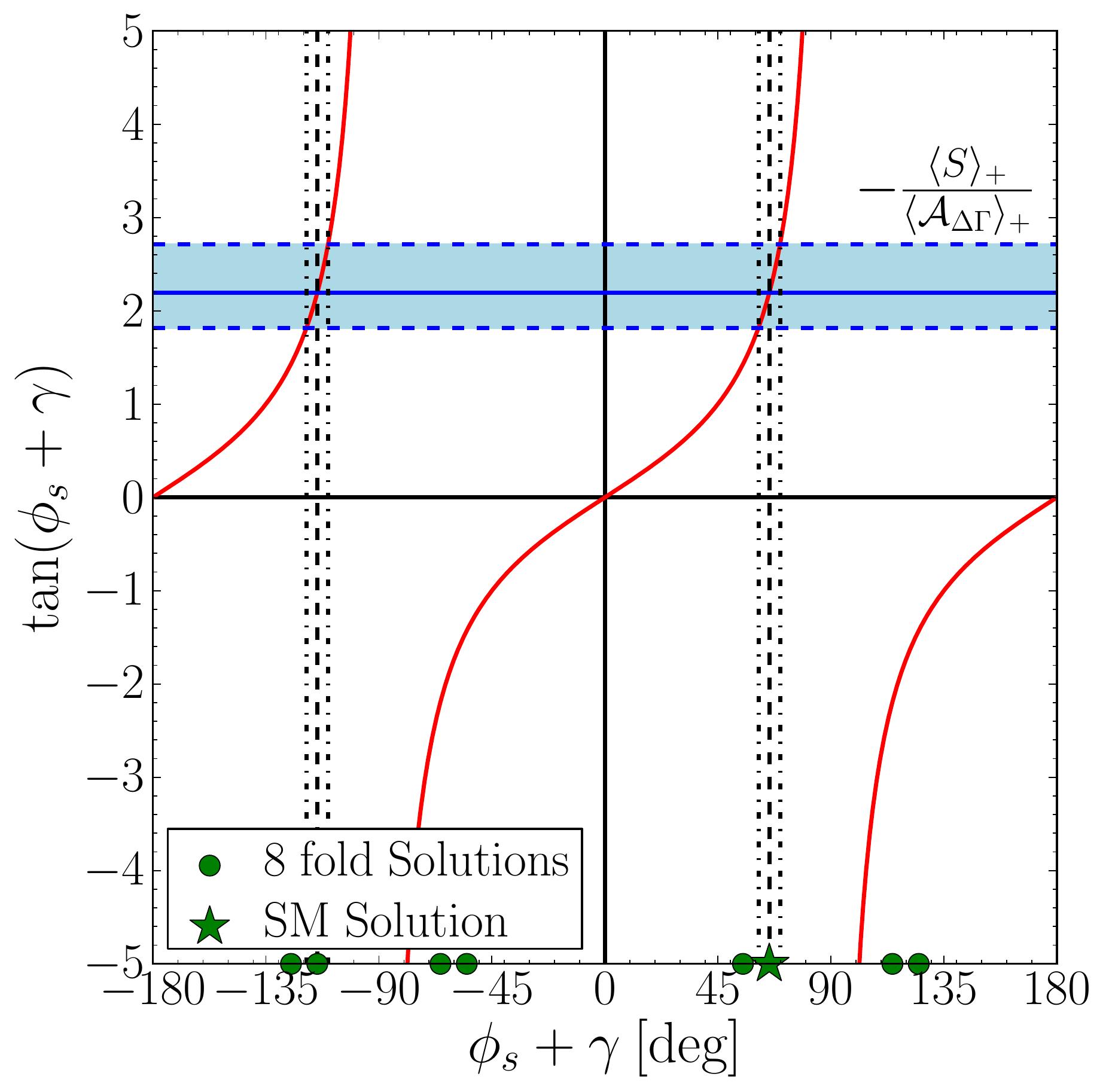}
\caption{Illustration of the determination of $\phi_s+\gamma$ from 
$\langle {\cal A}_{\Delta\Gamma}\rangle_+$ and $\langle S\rangle_+$ by means
of (\ref{extractPhiW}). We also show the eightfold solution 
resulting from the method (\ref{conv}) that does not use untagged information,
as discussed in the text.
The increasing experimental sensitivity of the panels from left to right corresponds 
to expectations of the LHCb experiment by the end of 2012, before the upgrade and 
after the upgrade, respectively.}\label{fig:toy_rel}
\end{figure}

Experimental data sets are simulated assuming approximate detector performance,
as discussed in Ref.~\cite{LHCb-CONF-2011-050-deltams} by the LHCb
collaboration, corresponding to a decay-time resolution of 50\,fs, a flavour
tagging efficiency of 38\%, and a wrong-tag probability of 34\%.
The sensitivity
is estimated for data sets that would correspond to about 1100 events per
fb$^{-1}$ of collected integrated luminosity~\cite{LHCb-BsDsK}, selecting only $B_s$
candidates with a lifetime of $t>0.5$~ps. Systematic effects, such as the presence of
background events, are ignored in this study.

In the toy simulation, the observables listed in (\ref{eq:theo_pred}) are determined from a fit to the
decay distributions from 3500 simulated $B_s\rightarrow D_s^\pm K^\mp$ events
corresponding to the approximate data sample that can be collected by the LHCb
experiment by the end of 2012.  The fit is repeated for 2000 different data
sets, resulting in an estimate for the sensitivity for the observables, which is
comparable to the accuracy of the prediction itself.  
In Table \ref{Tab:toy}, the statistical uncertainties for the observables are listed for 
data samples corresponding to the expected integrated luminosity of the LHCb experiment 
at the end of 2012, before the upgrade, and after the upgrade.
In our toy simulations an average correlation of 0.5 was observed between 
the $\mathcal{A}_{\Delta\Gamma}$ and $\overline{\mathcal{A}}_{\Delta\Gamma}$ 
observables, which is taken into account in the fits below; the correlations 
between the other CP observables is found to be negligible.

As a final step, these estimated experimental uncertainties for the observables 
$\mathcal{A}_{\Delta\Gamma}$, $S$, $C$ and their CP conjugates
can be translated into a determination
for $\gamma$. Using only  $S$, $\overline{S}$ and $C$, $\overline{C}$ 
following the 
approach without using untagged information, as described by (\ref{conv}), 
the experimental sensitivity is not sufficient 
to determine $\gamma$ for a data sample of about 3500 events, which  
can be collected by the end of 2012. A factor five increase in data size, 
corresponding to the end of the current LHCb experiment, would result in a
sensitivity of $\gamma+\phi_s = (65.6^{+17}_{-26})^\circ$
if the solution around the input value for $\gamma$ is selected.

\begin{figure}
\center
\includegraphics[width=0.32\textwidth]{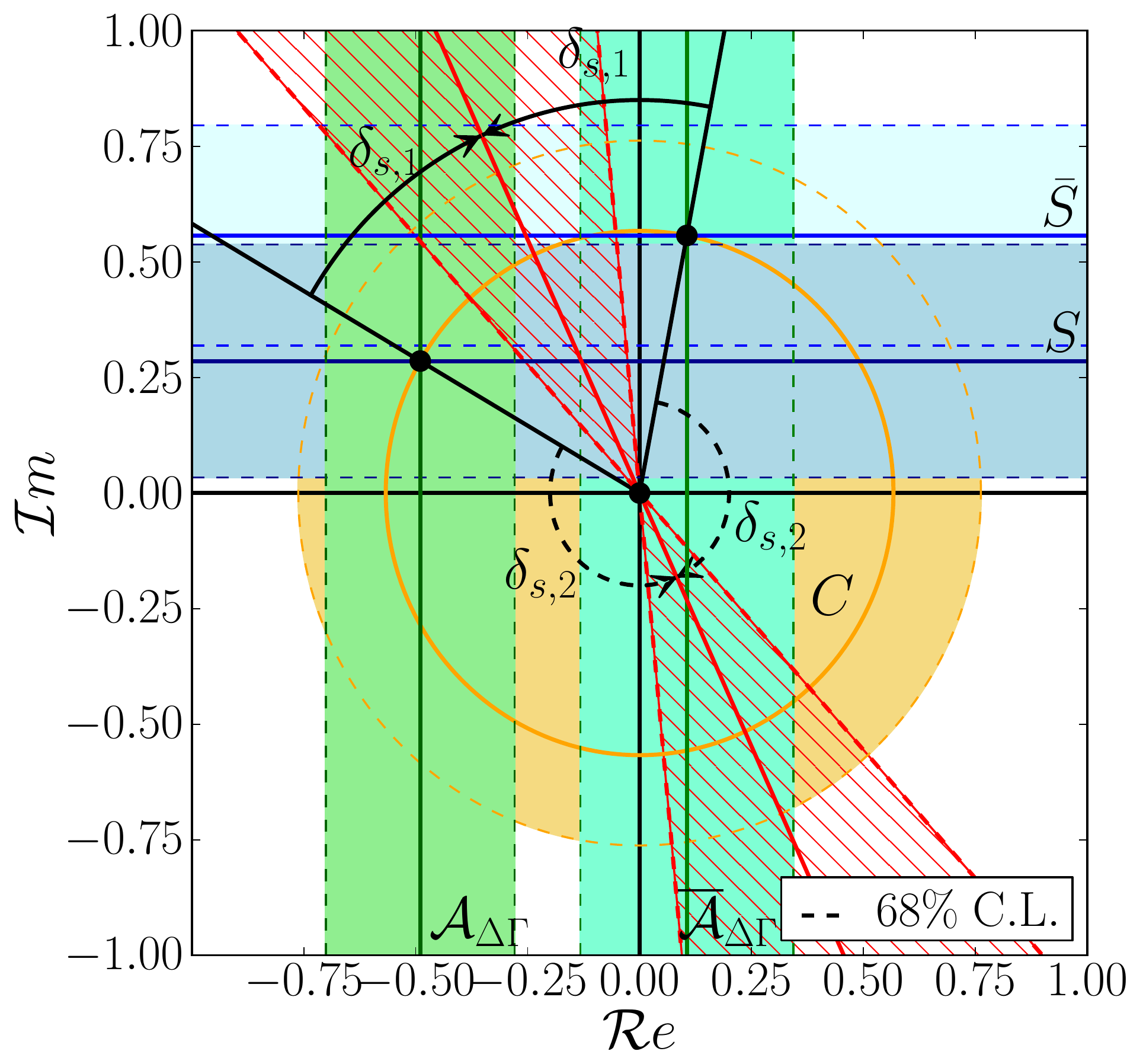}
\includegraphics[width=0.32\textwidth]{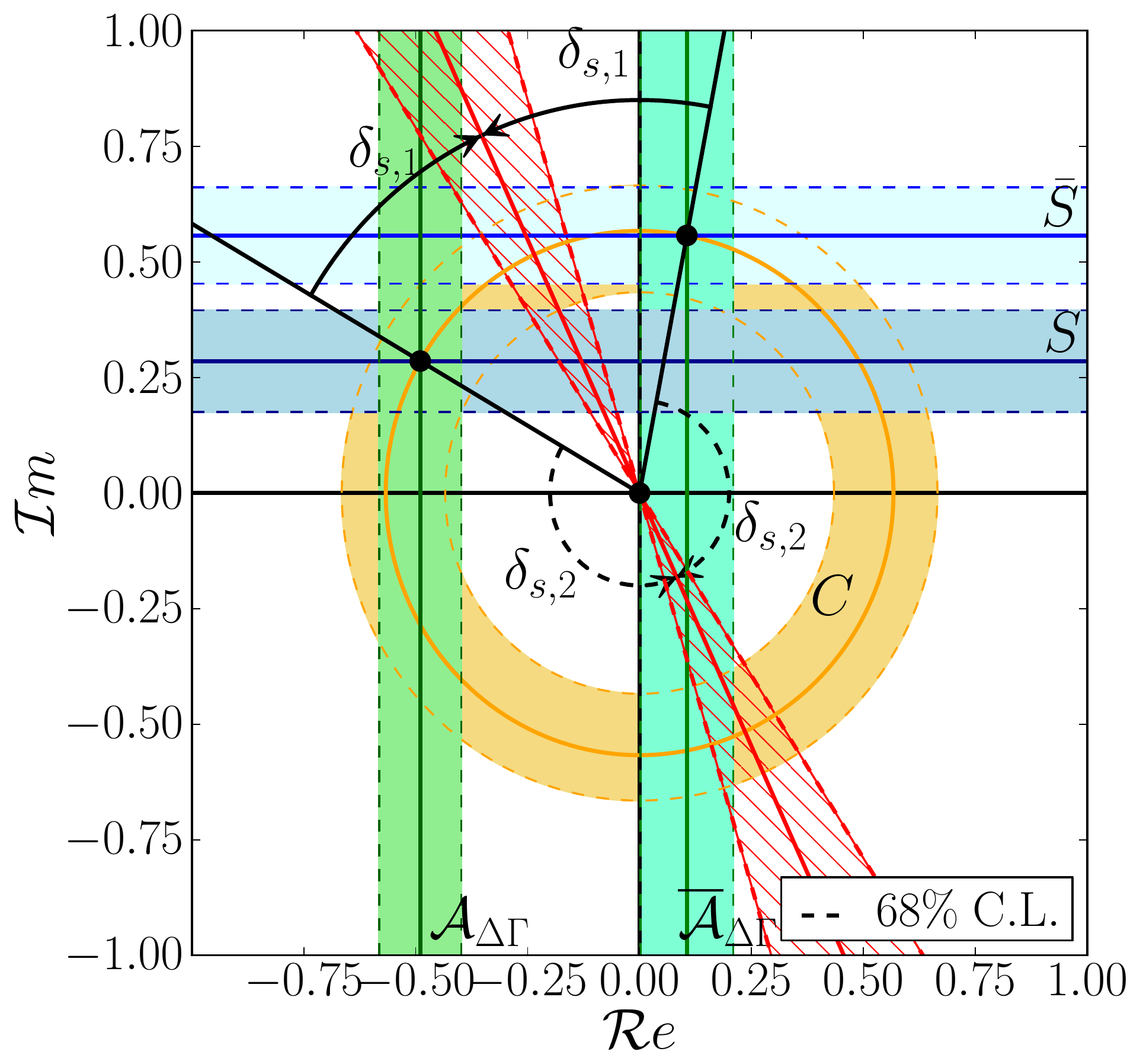}
\includegraphics[width=0.32\textwidth]{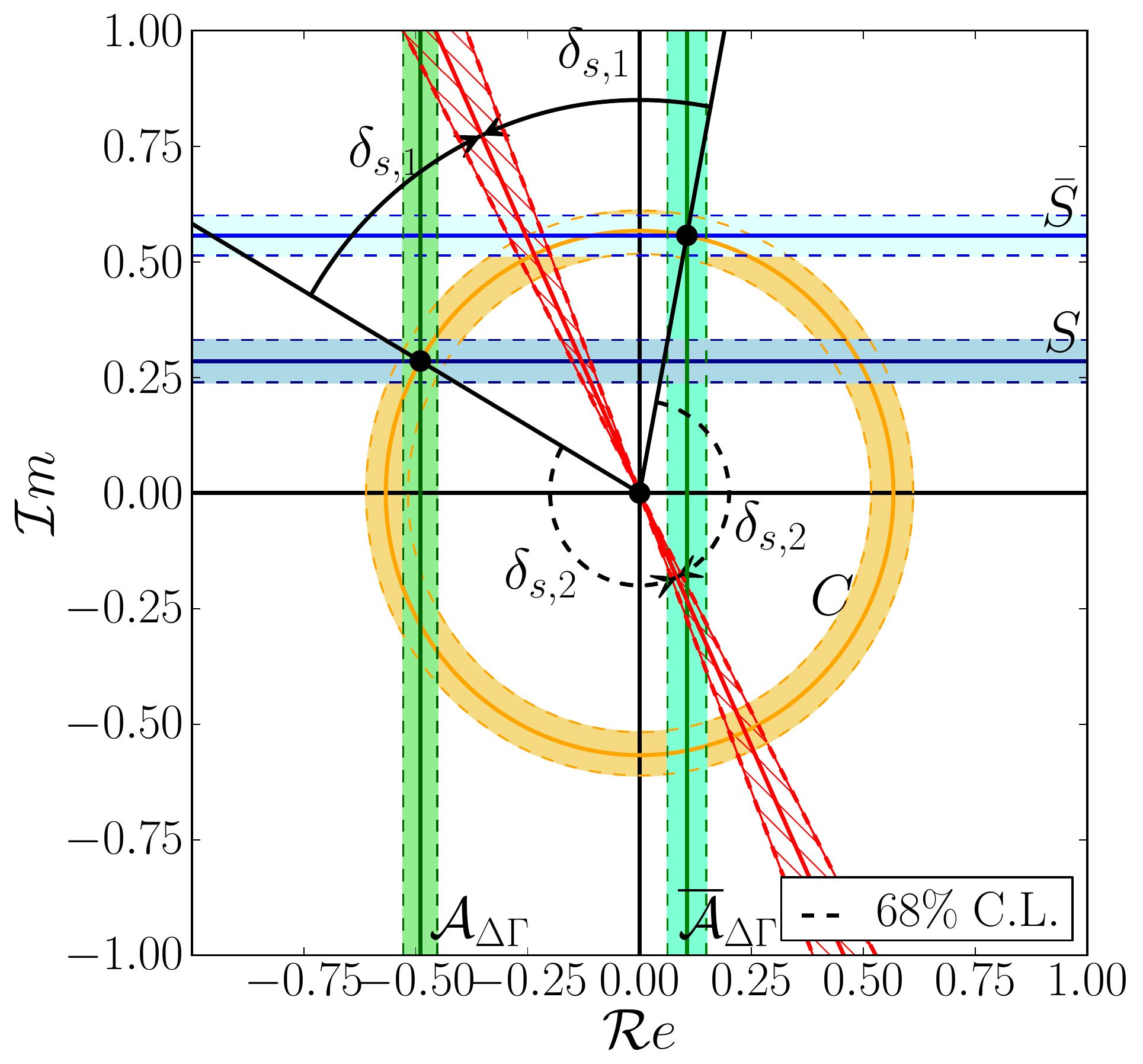}
\caption{Illustration of the determination of $\phi_s+\gamma$ from a simultaneous 
fit to ${\cal A}_{\Delta\Gamma}$, ${\cal S}$, $C$ and their CP conjugates in the complex plane
(see also Fig.~\ref{fig:polar}). 
The 68\% confidence levels for $\phi_s+\gamma$ and the CP observables are 
indicated by the hatched and shaded regions, respectively.
The two solutions shown for  $\phi_s+\gamma$ and $\delta_s$ 
correspond to the remaining twofold discrete ambiguity discussed in the text.
The increasing experimental sensitivity of the panels from left to right corresponds 
to expectations of the LHCb experiment by the end of 2012, before the upgrade and 
after the upgrade, respectively.}
\label{fig:toy_CPobs}
\end{figure}

If instead the observable pairs
\begin{equation}\label{comp-1}
\langle {\cal A}_{\Delta\Gamma} \rangle_+ + i\langle S \rangle_+=
-(-1)^L\sqrt{1-\langle C\rangle^2_-}\cos\delta_s \, e^{-i(\phi_s+\gamma)}
\end{equation}
and
\begin{equation}\label{comp-2}
\langle {\cal A}_{\Delta\Gamma} \rangle_- + i\langle S \rangle_-=
-(-1)^L\sqrt{1-\langle C\rangle^2_-}\sin\delta_s \, e^{i(\pi/2-(\phi_s+\gamma))}
\end{equation}
are used separately, the 2012 data sample corresponds to experimental 
sensitivities for $\gamma+\phi_s$ of $\pm 24^\circ$ and $\pm 29^\circ$, respectively; 
see the left panel of Fig.~\ref{fig:toy_angle}. In Fig.~\ref{fig:toy_rel}, we illustrate the
extraction of $\gamma+\phi_s$ from $\langle S\rangle_+$ and 
$\langle{\cal A}_{\Delta\Gamma}\rangle_+$ by means of the first relation 
in (\ref{extractPhiW}).
Finally, combining all the observables, i.e.\ ${\cal A}_{\Delta\Gamma}$, ${\cal S}$ 
and $C$ with their CP conjugates,
improves the sensitivity to $\pm 17^\circ$, as illustrated in the left panel 
of Fig.~\ref{fig:toy_CPobs}.  

With increasing data samples, shown in the 
middle and right panels of 
Figs.~\ref{fig:toy_angle}--\ref{fig:toy_CPobs}, the precision on 
the measurement of $\gamma+\phi_s$ is expected to increase
to about $\pm 7^\circ (3^\circ)$ using 18k(130k) 
$B_s\rightarrow D_s^\pm K^\mp$ events, which could finally be collected by the
current (upgraded) LHCb experiment, assuming
unchanged trigger and tagging performance. The sensitivity quoted here is different, but compatible 
with the projected sensitivity quoted by the LHCb collaboration~\cite{LHCb-upgrade}, 
where more sophisticated estimates are made for the trigger performance in the coming 
running periods. The different experimental errors
for the determination of $\phi_s+\gamma$, $\delta_s$ and $x_s$ from the 
$B_s\rightarrow D_s^{\pm}K^{\mp}$
decays are collected in Table~\ref{Tab:toy2}.

The magnitude of $\mathcal{A}_{\Delta\Gamma}+iS$
can be further constrained  through the $SU(3)$ flavour symmetry, i.e.\
through (\ref{xs-BR-det}) or by means of (\ref{eq:xsxd}) with (\ref{eq:xdsquared}). However, 
we find that this input, which would introduce the $SU(3)$ flavour symmetry into a theoretically
clean strategy, does not significantly improve the precision for $\gamma+\phi_s$.

On the other hand, if also decays of the type $B_s\rightarrow D_s^{*\pm}K^{\mp}$ 
can be reconstructed, the precision could be further enhanced in a theoretically clean way.
These channels require the reconstruction of a radiative photon in the decay 
$D_s^{*\pm}\rightarrow D_s^\pm \gamma$ and as such are  
experimentally more challenging. In Ref.~\cite{ThesisTilburg}, 
a gain in statistics of 28\% is deemed possible,
leading to an improvement of 13\% on the determination of  $\phi_s + \gamma$.

\boldmath
\section{Conclusions}\label{sec:concl}
\unboldmath
The decays $B_s \to D_s^{(*)\pm} K^\mp$ offer an interesting playground for the
LHCb experiment in this decade. We have performed a detailed analysis of the 
observables of these channels, addressing in particular the impact of the sizable $B_s$
decay width difference $\Delta\Gamma_s$, which has recently been established. 
This quantity leads to a subtle difference between the experimental and theoretical
branching ratios of the $B_s \to D_s^{(*)\pm} K^\mp$ decays, which can be resolved
experimentally through time information on the corresponding untagged data samples,
such as measurements of the effective decay lifetimes. 
We derived a lower bound for the ratio of the experimental $B_s\to D_s^\pm K^\mp$ and 
$B_s\to D_s^\pm \pi^\mp$ branching ratios given in (\ref{LHCb-CDF-BR}),
and observe that the central value for the LHCb result is too small by about two standard deviations.

The width difference $\Delta\Gamma_s$ offers the untagged observables 
${\cal A}_{\Delta\Gamma}$ and $\overline{\cal A}_{\Delta\Gamma}$ for the
final states $D_s^{(*)+}K^-$ and $D_s^{(*)-}K^+$,  respectively, which can nicely 
be combined with the corresponding mixing-induced CP asymmetries $S$ 
and $\overline{S}$ to determine $\phi_s+\gamma$ in an unambiguous way. 
We have illustrated this strategy and have obtained predictions for the 
$B_s \to D_s^{(*)\pm} K^\mp$ observables from an $SU(3)$ analysis of the 
$B$-factory data for $B_d\to D^{(*)\pm}\pi^\mp$,  $B_d\to D^{\pm}_s\pi^\mp$
decays. Moreover, making experimental simulations, we have shown that the 
interplay between the untagged observables 
${\cal A}_{\Delta\Gamma}$, $\overline{\cal A}_{\Delta\Gamma}$ and the tagged
CP asymmetries $S$, $\overline{S}$ is actually the key feature for being
able to measure $\phi_s+\gamma$ through the $B_s \to D_s^{(*)\pm} K^\mp$ decays
at LHCb. In this sense, the favourably large value of $\Delta\Gamma_s$ is a present
from Nature. 

\vspace*{1.0truecm}

{\bf 
\noindent Acknowledgements}

\vspace*{0.2truecm}

\noindent
This work is supported by the Netherlands Organisation for Scientific
Research (NWO) and the Foundation for Fundamental Research on Matter (FOM).
We thank Suvayu Ali for useful discussions.

\end{document}